\documentclass[a4paper,11pt]{article}
\pdfoutput=1 

\usepackage{jheppub} 
\usepackage[usenames,dvipsnames,table]{xcolor}
\usepackage[T1]{fontenc} 

\usepackage{multirow}
\usepackage{makecell}
\usepackage{mathrsfs}
\usepackage{amsmath,amsthm,amssymb}
\usepackage{comment}
\usepackage{float}
\numberwithin{equation}{section}
\usepackage{lastpage} 

\usepackage{relsize}

\usepackage{listings}
\usepackage{xcolor}
\newcommand{\bec}{\begin{center}}
\newcommand{\eec}{\end{center}}
\newcommand{\beq}{\begin{equation}}
\newcommand{\eeq}{\end{equation}}
\newcommand{\bea}{\begin{eqnarray}}
\newcommand{\eea}{\end{eqnarray}}
\newcommand{\nn}{\nonumber}

\newcommand{\Tr}{{\rm Tr~}}
\newcommand{\hf}{\frac{1}{2}}
\newcommand{\qtr}{\frac{1}{4}}
\newcommand{\psib}{{\overline{\psi}}}
\newcommand{\B}{{\mathcal{B}}}
\newcommand{\Q}{{\mathcal{Q}}}
\newcommand{\Qb}{{\overline{\mathcal{Q}}}}
\newcommand{\mcS}{\mathcal{S}}

\newcommand{\cD}{{\cal D}}

\newcommand{\cN}{{\cal N}}
\newcommand{\cO}{{\cal O}}

\usepackage[caption=false]{subfig}

\title{\boldmath Complex Langevin Dynamics and Supersymmetric Quantum Mechanics}

\author[]{Anosh Joseph and}
\author[]{Arpith Kumar}

\affiliation[]{Department of Physical Sciences,\\ Indian Institute of Science Education and Research (IISER) Mohali, \\Knowledge City, Sector 81, SAS Nagar, Punjab 140306, India}

\emailAdd{anoshjoseph@iisermohali.ac.in}
\emailAdd{arpithk.iiserm@gmail.com}

\abstract{Using complex Langevin method we probe the possibility of dynamical supersymmetry breaking in supersymmetric quantum mechanics models with complex actions. The models we consider are invariant under the combined operation of parity and time reversal, in addition to supersymmetry. When actions are complex traditional Monte Carlo methods based on importance sampling fail. Models with dynamically broken supersymmetry can exhibit sign problem due to the vanishing of the partition function. Complex Langevin method can successfully evade the sign problem. Our simulations suggest that complex Langevin method can reliably predict the absence or presence of dynamical supersymmetry breaking in these one-dimensional models with complex actions.
}

\arxivnumber{2011.08107}

\begin{document} 

\maketitle
\flushbottom

\section{Introduction}

The phenomenon of spontaneous breaking of supersymmetry (SUSY) requires an understanding of non-perturbative aspects of quantum field theory \cite{Witten:1981nf, Witten:1982df}. Lattice regularized form of the field theory path integral provides a systematic tool to investigate numerous non-perturbative features of quantum field theories. Monte Carlo methods have been used to reliably extract the physics of such systems. The basic idea behind path integral Monte Carlo is to generate field configurations with a positive definite probability weight, given by the exponential of the negative of the action (in Euclidean spacetime), and then compute the path integral by statistically averaging these importance sampled ensemble of field configurations. However, when the action is complex, for example, when studying QCD at finite temperature and baryon/quark chemical potential, QCD with a theta term, Chern-Simons gauge theories, or chiral gauge theories, the fermion determinant of the theory can be complex, and this feature results in the so-called {\it sign problem}. This makes the simulation algorithms based on path integral Monte Carlo unreliable. 

There exist very few methods in the literature that can effectively handle or evade this infamous sign problem. Some of these include methods such as analytical continuation, Taylor series expansion \cite{deForcrand:2002hgr}, methods based on the complexification of the integration variables such as the Lefschetz thimble method \cite{PhysRevD.86.0745006} and the complex Langevin (CL) method \cite{Klauder:1983nn, Klauder:1983zm, Klauder:1983sp, Parisi:1984cs, Damgaard:1987rr}. (See Refs. \cite{Cristoforetti:2012su, Fujii:2013sra, DiRenzo:2015foa, Tanizaki:2015rda, Fujii:2015vha, Alexandru:2015xva} for simulations based on the Lefschetz thimble method.) The CL method is based on stochastic quantization, and it is a straightforward generalization of the real Langevin method into complexified field configurations. 

The central theme of stochastic quantization is that the expectation values of observables are obtained as the equilibrium values of a stochastic process. In Langevin dynamics, this is achieved by evolving the system in a fictitious time direction $\theta$ (Langevin time) subject to Gaussian stochastic noise. The idea of stochastic quantization for ordinary field theory systems with real actions is extended to the cases with complex actions, and this overcomes the sign problem by defining a stochastic process, with complexified field variables, using Langevin equations for the complex action \cite{Klauder:1985kq, Klauder:1985ks,Gausterer:1986gk}. The use of Langevin equations, rather than importance sampling, waives the restriction of real and positive semi-definite measures. Such a strategy makes the real field variables $\phi$ complex, during the Langevin evolution, since the gradient of the action, the {\it drift term}, becomes complex. 

The complex Langevin equation in Euler discretized form for the field variable $\phi$ reads
\beq
\phi (\theta + \epsilon) = \phi (\theta) - \left[ \frac{\delta S[\phi]}{\delta \phi (\theta)} \right] \epsilon +  \eta (\theta) \sqrt{\epsilon},
\eeq
where $S[\phi]$ is the action of the theory, the Langevin time is discretized as an integer multiple of the Langevin step size $\epsilon$, and $\eta (\theta)$ is a Gaussian noise satisfying the constraints $\langle \eta (\theta) \rangle = 0$ and $\langle \eta (\theta) \eta (\theta ') \rangle = 2 \delta_{\theta \theta  '}$. In our simulations, we use real Gaussian stochastic noise to control excursions in the imaginary directions of the field configurations \cite{Aarts:2011ax}.

Noise averaged expectation value for an arbitrary operator $\cO[\phi]$ can be defined as
\beq
\left \langle \cO[\phi(\theta)] \right \rangle_\eta = \int d\phi P[\phi(\theta)] \cO[\phi],
\eeq
where the probability distribution $P[\phi(\theta)]$ satisfies the Fokker-Planck equation
\beq
\frac{\partial P[\phi(\theta)]}{\partial \theta} = \frac{\delta}{\delta \phi(\theta)} \left[ \frac{\delta}{\delta \phi(\theta)} + \frac{\delta S[\phi]}{\delta \phi(\theta)} \right] P[\phi(\theta)].
\eeq

In the case of real action it can be easily shown that in the large Langevin time limit, $\theta \to \infty$, the stationary solution of the Fokker-Planck equation
\beq
P[\phi] \sim \exp \left( - S[\phi] \right)
\eeq
will be reached, and thus ensuring the convergence of the Langevin dynamics to the correct equilibrium distribution. When the action is complex, the situation is challenging. The field configurations are complexified due to the drift term being a complex quantity and thus, we end up with complex probabilities. The CL method became very popular when it was first proposed in the 1980s, but certain problems were encountered shortly after. First was the problem of runaways, where the field configurations would not converge, and the second was the convergence to a wrong limit. In recent years, the CL method has been successfully revived, producing correct results even when the {\rm sign problem} is severe. When applying the CL method, it is essential that the expectation values computed with a real probability $P[\phi]$ along the complex trajectory $\phi = x+ iy$ are in agreement with the original expectation values with complex weight $\rho(\phi) = e^{-S(\phi)}$. That is,
\beq
\left \langle \cO \right \rangle = \int dx dy P[x, y] \cO[x + iy] = \int d \phi \rho[\phi] \cO[\phi].
\eeq

Recently, the proof towards convergence to the complex weight in the above equivalence has been an object of thorough study, and correctness criteria have been proposed for the validity of the method \cite{Aarts:2009uq, Nagata:2015uga}. In Appendix. \ref{subsec:reliability}, we discuss these criteria and investigate the reliability of the simulations carried out in this work. The revived interest in this field has led to promising developments in the optimization of the method. It has been noted that the complexification of field variables can drastically influence the Langevin trajectory, making large excursions in the imaginary direction. For instance, it might get closer to unstable directions. An efficient algorithm to evade such a situation involves considering adaptive Langevin step size in the numerical integration of Langevin equations \cite{Aarts:2009dg}. Ref. \cite{Damgaard:1987rr} provides a pedagogical review on this method, and Ref. \cite{Berger:2019odf} is a recent review in the context of the sign problem in quantum many-body physics. 

The CL method has been used successfully in the  study of various models in the recent past \cite{Berges:2005yt, Berges:2006xc, Berges:2007nr, Bloch:2017sex, Aarts:2008rr, Pehlevan:2007eq, Aarts:2008wh, Aarts:2009hn, Aarts:2010gr, Aarts:2011zn}. There have also been studies of spontaneous symmetry breaking (SSB) induced by complex fermion determinant \cite{Ito:2016efb, Ito:2016hlj, Anagnostopoulos:2017gos} and a recent study addressing the SSB of $SO(D)$ rotational symmetry based on CL method \cite{Anagnostopoulos:2020xai}. In Ref. \cite{Basu:2018dtm} the authors used CL simulations to observe the Gross-Witten-Wadia (GWW) \cite{Gross:1980he, Wadia:2012fr, Wadia:1980cp} phase transition in certain large-$N$ matrix models. In Ref. \cite{Joseph:2019sof} the authors looked at certain classes of zero-dimensional supersymmetric quantum field theories with complex actions using the CL method. In this work we will investigate dynamical SUSY breaking in an interesting class of supersymmetric actions that are complex and exhibiting invariance under the combine operation of parity and time reversal  (${\cal PT}$) symmetry. This class of actions cannot be studied using traditional Monte Carlo methods.

The paper is organized as follows. In Sec. \ref{cont-susy-mechanics} we briefly introduce supersymmetric quantum mechanics with two supercharges. In Sec. \ref{lattice-susy-theory} we discuss the lattice regularization of the model. There, we also provide a set of useful observables that can probe SUSY breaking in the system. They include correlation functions and Ward identities. In Sec. \ref{lattice-simulation} we present the simulation results of various models including the ${\cal PT}$ symmetric model. Conclusions are provided in Sec. \ref{sec:Conclusions}. In Appendix \ref{app:FP-correctness} we study the reliability of simulations using the Langevin operator and in Appendix \ref{app:drift-decay} we examine the correctness of simulations by observing the fall-off behavior of the probability distributions of the drift term magnitudes. 

\section{Supersymmetric quantum mechanics}
\label{cont-susy-mechanics}

Let us consider the action $S[\phi, \psi, \psib]$ of a supersymmetric quantum mechanics with a general superpotential $W(\phi)$. The model is invariant under two supercharges. The degrees of freedom are a scalar field $\phi$ and two fermions $\psi$ and $\psib$. We take the action to be an integral over a compactified time circle of circumference $\beta$ in Euclidean time. It has the form
\bea
\label{cont-action-1d}
S[\phi, \psi, \psib] &=& \int_0^\beta d \tau
\Bigg[ \hf \B(\tau)^2 + i \B \left( \frac{\partial}{\partial{\tau}}{\phi(\tau)} + \frac{\partial }{\partial{\phi}} W(\phi(\tau)) \right)  \nn \\ 
&& \hspace{2cm} + {\psib(\tau)} \left(\frac{\partial}{\partial{\tau}} + \frac{\partial^2 }{\partial{\phi}^2} {W} (\phi(\tau)) \right) \psi(\tau) \Bigg],
\eea
where $\B$ is an auxiliary field and the derivatives with respect to $\tau$ and $\phi$ are denoted by a dot and a prime, respectively.

Denoting the two supercharges by $\Q$ and $\Qb$, the SUSY transformations have the form 
\beq
\label{Q-transf}
\Q \phi = \psi,~~~ \Q \psi = 0,~~~ \Q \psib =- i \B,~~~ \Q \B = 0,
\eeq
and
\beq
\label{Qb-transf}
\Qb \phi = - \psib,~~~ \Qb \bar{\psi} = 0,~~~ \Qb \psi = - i \B + 2 \dot{\phi},~~~ \Qb \B = 2 i \dot{\psib}. 
\eeq

The supercharges satisfy the algebra
\beq
\label{eq:algebra}
\{ \Q, \Q \} = 0, \quad \{ \Qb, \Qb \} = 0, \quad \{ \Q, \Qb \} = 2\partial_{\tau}.
\eeq

We also note that the action can be expressed in $\Q$- and $\Q \Qb$-exact forms. That is,
\beq
\label{QQb-cont-exact}
S = \Q  \int_0^\beta d{\tau} ~ \psib \left[ \frac{i}{2} \B - \left(\frac{\partial \phi}{ \partial \tau} + W^{'}(\phi)  \right) \right] = \Q \Qb  \int_0^\beta d\tau \left( \hf \psib \psi + W (\phi) \right).
\eeq

The partition function in path integral formalism is defined by
\beq
\label{eqn:cont-pf}
Z \equiv \int \cD \B \cD \phi \cD \psi  \cD \psib ~e^{-S[\phi, \psi, \psib]}
\eeq
with periodic temporal boundary conditions for all the fields. 

In a system with unbroken SUSY, we can consider a Hamiltonian $H$ corresponding to the Lagrangian in Eq. \eqref{cont-action-1d} having energy levels $E_n$ where $n = 0, 1, 2, \dots$, such that the ground state energy $E_0=0$. Then, the bosonic and fermionic excited states form a SUSY multiplet
\beq
\label{eqn:susy_multiplet_unbroken}
|b_{n+1} \rangle = \frac{1}{\sqrt{2 E_{n+1}}} \Qb |f_n \rangle, ~~~|f_n \rangle = \frac{1}{\sqrt{2 E_{n+1}}} \Q |b_{n+1} \rangle
\eeq
satisfying the algebra given in Eq. \eqref{eq:algebra}, with $|b_{0} \rangle$ being the ground state of the system. Assuming that the states $|b_n \rangle$ and $|f_n \rangle$ have the fermion number charges $F = 0$ and $F = 1$, respectively, when periodic temporal boundary conditions are imposed for both the bosonic and fermionic fields, $Z$ is equivalent to the Witten index $\Delta_W \left( \beta\right)$ \cite{Witten:1982df}.  It is easy to see that the partition function defined as
\beq
\label{eqn:cont-Witten_index_unbroken}
Z  = \Delta_W \left(\beta \right) = \Tr \left[ (-1)^F e^{-\beta H} \right] 
=  \langle b_{0} | b_{0} \rangle  +\sum_{n = 0}^\infty \left[ \left(\langle b_{n+1} | b_{n+1} \rangle - \langle f_n | f_n \rangle  \right) e^{-\beta E_{n+1}} \right] 
\eeq
does not vanish due to the existence of a normalizable ground state. As a result, the normalized expectation values of observables are well-defined. 

The normalized expectation value of the auxiliary field defined as
\beq
\label{eqn:aux_field_exp_val}
\langle \B \rangle = \frac{1}{\Delta_W \left(\beta \right)} \left[\langle b_{0}|\B | b_{0} \rangle    +  \sum_{n = 0}^\infty \left(\langle b_{n+1}|\B | b_{n+1} \rangle - \langle f_n | \B |f_n \rangle \right) e^{-\beta E_{n+1}} \right]
\eeq
can be used as an order parameter to probe SUSY breaking \cite{Kuroki:2009yg}. (We note that the unpaired state appearing in Eqs. \eqref{eqn:cont-Witten_index_unbroken} and \eqref{eqn:aux_field_exp_val} need not have to be a bosonic state or a unique state.) The auxiliary field was introduced for the off-shell completion of the SUSY algebra. The $\Q$ transformation of $\B$ in Eq. \eqref{Q-transf} and the fact that the ground state is annihilated by the supercharges together imply that in the absence of SUSY breaking the normalized expectation value of the auxiliary field vanishes. However, in the SUSY broken case, we end up in a not-so-trivial situation.

In a system with SUSY spontaneously broken, the Hamiltonian $H$ corresponding to the Lagrangian in Eq. \eqref{cont-action-1d} has a positive ground state energy ($0 < E_0 < E_1 < E_2 \dots$), and the SUSY multiplet is defined as
\beq
\label{eqn:susy_multiplet}
|b_n \rangle = \frac{1}{\sqrt{2 E_n}} \Qb |f_n \rangle, ~~~|f_n \rangle = \frac{1}{\sqrt{2 E_n}} \Q |b_n \rangle,
\eeq
satisfying the algebra given in Eq. \eqref{eq:algebra}. Differently from the unbroken SUSY case, when SUSY is broken, the supersymmetric partition function
\beq
\label{eqn:cont-Witten_index}
Z  = \Delta_W = \Tr \left[ (-1)^F e^{-\beta H} \right] = \sum_{n = 0}^\infty \left[ \left(\langle b_n | b_n \rangle - \langle f_n | f_n \rangle  \right) e^{-\beta E_n} \right]
\eeq
vanishes due to the cancellation between bosonic and fermionic states.  As a consequence, the normalized expectation values of observables will be ill-defined. We consider the auxiliary field as an observable, and the normalized expectation value can be computed as
\beq
\label{eqn:cont-B-expectation}
\langle \B \rangle = \frac{1}{\Delta_W \left(\beta \right)} \sum_{n = 0}^\infty \left[ \left(\langle b_n|\B | b_n \rangle - \langle f_n | \B |f_n \rangle\right) e^{-\beta E_n} \right],
\eeq
where the numerator vanishes from $\Q$-supersymmetry $ \left( \langle b_n|\B | b_n \rangle = \langle f_n | \B |f_n \rangle \right)$. Thus, in a system with broken SUSY, the normalized expectation of auxiliary field admit a $0/0$ indefinite form.

In Ref. \cite{Kuroki:2009yg} Kuroki and Sugino introduced a regulator that explicitly breaks SUSY and resolves the degeneracy by fixing a single vacuum state in which SUSY is broken. This regulator $\alpha$ (the twist parameter) can be implemented by imposing twisted boundary conditions (TBC) for fermions. That is, 
\beq
\psi (\tau + \beta) = e^{i \alpha} \psi(\tau) \quad {\rm and} \quad \psib(\tau +\beta) = e^{-i \alpha} \psib(\tau).
\eeq
It was shown in Ref. \cite{Kuroki:2009yg} that for a non-zero $\alpha$, the partition function does not vanish, and the normalized expectation value of the auxiliary field is well-defined. In the limit $\alpha \to 0$, PBCs are recovered and supersymmetry is restored. Thus $\alpha$ regularizes the indefinite form in Eq. \eqref{eqn:cont-B-expectation}. Now, vanishing expectation value of the auxiliary field in the limit $\alpha \to 0$ suggests that SUSY is not broken, while a non-zero value suggests that SUSY is broken. We incorporate the twist $\alpha$ when we introduce the lattice regularized theory in Sec. \ref{sec:lattice-theory}. 

It is possible to integrate out the auxiliary field using its equation of motion
\beq
\B = - i \left( \frac{\partial \phi}{ \partial \tau} + W^{'}(\phi) \right),
\eeq
to get the on-shell form of the action
\beq
\label{eqn:act-cont}
S = \int_0^\beta d\tau
\left[ \hf \left\{ {\left(\frac{\partial \phi}{ \partial \tau}\right)} + W^{'}(\phi) \right\}^2 + {\psib} \left\{ \frac{\partial }{ \partial \tau} + W''(\phi) \right\} \psi \right].
\eeq

Upon using the Leibniz integral rule and discarding the resultant total derivative term, the action takes the form
\beq
S = \int_0^\beta d\tau
\left[ \hf \left\{ {\left(\frac{\partial \phi}{ \partial \tau}\right)}^2 + \left[ W'(\phi) \right]^2 \right\} + {\psib} \left\{ \frac{\partial }{ \partial \tau} + W''(\phi) \right\}\psi \right].
\eeq
In the above expression, the total derivative term we omitted was $(\partial \phi / \partial \tau) W'(\phi)$. Note that such an omission is only possible in the continuum theory. When we discretize the theory on a lattice, this term does not vanish, and its presence is crucial to ensure the $\Q$-exact lattice supersymmetry. Thus, in our lattice analysis, we will use Eq. \eqref{eqn:act-cont} as the continuum target theory.

\section{Lattice regularized models}
\label{lattice-susy-theory}

We discretize the action given in Eq. \eqref{eqn:act-cont} on a one-dimensional lattice. Let us take the lattice to be $\Lambda$, having $T$ number of equally spaced sites with lattice spacing $a$. The integral and continuum derivatives are replaced by a Riemann sum $a  {\Sigma}$ and a lattice difference operator $\nabla$, respectively. The physical extent of the lattice is defined as $\beta \equiv T a$. 

There are several ways to regularize a given theory on a lattice. We will choose the prescription in which the derivatives appearing in the action take the form of a symmetric difference operator
\beq
\nabla^S_{ij} = \hf \left( \nabla^+_{ij} + \nabla^-_{ij} \right),
\eeq
where
\bea
&& \nabla^+_{ij} = \frac{1}{a} (\delta_{i+1, j} - \delta_{i, j}) \quad \longrightarrow \quad \nabla^+_{ij} f_j = \frac{1}{a} ( f_{i+1} - f_i ), {\rm ~ and} \\ 
&& \nabla^-_{ij} = \frac{1}{a} (\delta_{i, j} - \delta_{i-1, j}) \quad \longrightarrow \quad \nabla^-_{ij} f_j = \frac{1}{a} (f_i - f_{i-1}), 
\eea
are the forward and backward difference operators, respectively, and $i, j$ represent lattice sites. However, it is known that the symmetric derivative leads to the so-called fermion doubling problem and this, in turn, leads to a non-supersymmetric lattice theory. We can use the Wilson discretization prescription to decouple these extra fermionic modes from the system. The difference operator is modified as
\beq
\nabla^W_{ij}(r) = \nabla^S_{ij} - \frac{r a}{2}\square_{ij},
\eeq
where $\square_{ij} = \nabla^+_{ik} \nabla^-_{kj}$ is the usual lattice Laplacian and the Wilson parameter $r \in \left[-1,1\right] / \left\{0 \right\}$ \cite{Baumgartner:2014nka}. For one-dimensional derivatives it turns out that the standard choice of $r = \pm 1$ yields $\nabla^W_{ij}(\pm 1) = \nabla^{\mp}_{ij}$, thereby suggesting that the doubling problem can be resolved by simply using forward or backward difference operator. The reason being that for any choice of the lattice difference operator, the theories can be made manifestly supersymmetric upon the addition of appropriate improvement terms corresponding to the discretization of continuum surface integrals \cite{Bergner:2007pu}. 

In our analysis, for the standard choice of the Wilson parameter, we follow the symmetric derivative with a Wilson mass matrix suggested in Ref. \cite{Catterall:2000rv}. The lattice regularized action then takes the form
\beq
\label{eqn:lat-reg-action}
\mcS = a \sum_{i = 0}^{T-1} \left[ \hf \left( \sum_{j = 0}^{T-1} \nabla^S_{ij} \phi_j + \Omega'_i  \right)^2 + \psib_i \sum_{j = 0}^{T-1} \left( \nabla^S_{ij} + \Omega''_{ij} \right) \psi_j \right],
\eeq
where the quantity $\Omega'_i$ is defined as 
\beq
\Omega'_i = \sum_{j = 0}^{T-1} K_{ij} \phi_j + W'_i,
\eeq
and its derivative $\Omega''_{ij} $ is $\Omega''_{ij} =  K_{ij} +  W''_{ij} \delta_{ij}$. The Wilson mass matrix $K_{ij}$ has the form $K_{ij} = m \delta_{ij} - \frac{r a}{2} \square_{ij}$.

We can make the variables dimensionless by performing appropriate rescaling. Let us consider the following set of redefinitions for the variables
\beq
\label{eqn:rescale}
\widetilde{\phi} = a^{-1/2} \phi, \quad \widetilde{\nabla}^S = a \nabla^S, \quad \widetilde{\Omega}' = \sqrt{a} \Omega', \quad \widetilde{\Omega}'' = a \Omega''.
\eeq

Under these rescalings the action becomes
\beq
\label{eqn:lat-dimless-action}
\widetilde{\mcS} = \sum_{i = 0}^{T-1} \left[ \hf \left( \sum_{j = 0}^{T-1} \widetilde{\nabla}^S_{ij} \widetilde{\phi}_j + \widetilde{\Omega}'_i \right)^2 + \psib_i \sum_{j = 0}^{T-1} \left( \widetilde{\nabla}^S_{ij} + \widetilde{\Omega}''_{ij}  \right) \psi_j\right].
\eeq

\subsection{Theory on a lattice}
\label{sec:lattice-theory}

For convenience, we will not be using the tilde sign on the dimensionless variables; all variables and fields mentioned from now on are understood to be dimensionless. Physical quantities will be labeled differently.

The supersymmetry transformations are modified to contain the Wilson mass terms. For a given lattice site $k$ they are given by  
\beq
\label{eqn:lat-Wilson-Q}
\Q \phi_k = \psi_k, \quad \Q \psib_k = - N_k, \quad \Q \psi_k = 0, 
\eeq
and
\beq
\label{eqn:lat-Wilson-Qbar}
\Qb \phi_k = - \psib_k, \quad \Qb \psi_k = \overline{N}_k, \quad \Qb \hspace{0.05cm} \psib_k = 0,
\eeq
where 
\beq
N_k = \nabla^S{\phi}_k + \Omega'_k, \quad {\rm and} \quad \overline{N}_k = \nabla^S{\phi}_k - \Omega'_k.
\eeq

The supercharges satisfy the algebra
\beq
\label{eqn:lat-Q-Qb-algebra}
\{ \Q, \Q \} = 0, \quad \{ \Qb, \Qb \} = 0,\quad{\rm and }\quad \{ \Q, \Qb \} = 2 \nabla^S.
\eeq

The main obstacle that prevents the preservation of exact lattice SUSY is the failure of the Leibniz rule for lattice derivatives. Unlike the continuum action given in Eq. \eqref{eqn:act-cont}, the lattice regularized action
\beq
\label{eqn:lat-action}
\mcS = \sum_{i = 0}^{T-1} \left[\hf \left( \sum_{j = 0}^{T-1} \nabla^S_{ij} \phi_j + \Omega'_i \right)^2 + \psib_i \sum_{j = 0}^{T-1} \left( \nabla^S_{ij} + \Omega''_{ij} \right) \psi_j \right]
\eeq
preserves only the $\Q$ supercharge. The $\Qb$ supersymmetry is broken for $T \geq 2$. It can also be shown that the action is only $\Q$ invariant. That is, $\Q \mcS = 0 \neq \Qb \mcS$.

The $\Qb$ supersymmetry is broken for finite lattice size $T$ because it is not possible to define a corresponding $\Qb$ invariant transformation on lattice variables such that the algebra $\{ \Q, \Qb \} = 2 \nabla^S$ still holds \cite{Kanamori:2007yx}. However, the $\Q$-exactness is essential and sufficient to kill any SUSY breaking counter-terms and thereby suppress lattice artifacts. See Refs. \cite{Catterall:2000rv, Catterall:2003wd, Giedt:2004vb, Kuroki:2009yg} for more discussions on this.

As mentioned in Sec. \ref{cont-susy-mechanics} for the continuum theory, when SUSY is broken, the partition function vanishes. In that case, the expectation values of the observables normalized by the partition function could be ill-defined. To overcome this difficulty in our lattice regularized theory, we will apply periodic boundary conditions for bosons and twisted boundary conditions for fermions \cite{Kuroki:2009yg, Kuroki:2010au}. 

Introducing the twist, we have $\phi_T = \phi_0, \quad \psi_T = e^{ i \alpha } \psi_0, \quad \psib_T = e^{ - i \alpha } \psib_0$. For the case of dynamically broken SUSY, when $\alpha = 0$, the Witten index vanishes, which dictates that the fermion determinant changes its sign depending on the boson field configurations. This is the sign problem in models with dynamical SUSY breaking.

The partition function given in Eq. \eqref{eqn:cont-pf} takes the following form 
\bea
\label{eqn:lat-pf}
Z_\alpha = \left( \frac{1}{\sqrt{2 \pi}} \right)^T  \int \left( \prod_{k = 0}^{T-1} d\phi_k d\psi_k d\psib_k \right) e^{ - \mcS_\alpha },
\eea
where $\mcS_\alpha$ is the lattice regularized action that respects the twisted boundary conditions. 

We have
\beq
\label{eqn:lat-action-r1}
\mcS_\alpha = \sum_{i = 0}^{T-1} \hf \bigg( \phi_i - \phi_{i-1} + m \phi_i + W'_i \bigg)^2 + \sum_{i = 0}^{T-1} \psib_i \bigg( \psi_i - \psi_{i-1} + \left( m + W''_{ii} \right) \psi_i \bigg).
\eeq

Let us absorb the mass term into the potential $W$, and define a new potential $\Xi$ as
\beq
\label{eqn:lat-anho-pot}
\Xi \equiv \hf m \phi^2 + W.
\eeq

The action with twisted boundary conditions now takes the form
\bea
\label{eqn:lat-action-r1-Xi}
\mcS_\alpha &=& \sum_{i = 0}^{T-1} \hf \bigg( \sum_{j = 0}^{T-1}  \nabla^{-}_{ij} \phi_j + \Xi'_i  \bigg)^2 + \sum_{i = 0}^{T-1} \psib_i  \bigg(\sum_{j = 0}^{T-1}  \nabla^{-}_{ij}  + \Xi^{''}_{ij} \bigg) \psi_j.  
\eea

Also the expressions for $N_i$ and $\overline{N}_i$ become
\bea
N_i = \sum_{j = 0}^{T-1} \nabla^S_{ij} \phi_j + \Omega'_i = \sum_{j = 0}^{T-1} \nabla^{-}_{ij} \phi_j + \Xi'_i, \\
\overline{N}_i =\sum_{j = 0}^{T-1} \nabla^S_{ij} \phi_j - \Omega'_i = \sum_{j = 0}^{T-1} \nabla^+_{ij} \phi_j - \Xi'_i. 
\eea

After integrating out fermions, the fermionic contribution to the partition function given in Eq. \eqref{eqn:lat-pf} has the form
\bea
\label{eqn:z-alpha-F}
Z_\alpha^F &=& \prod_{k = 0}^{T-1} \left(1+ \Xi^{''}_{kk} \right) - e^{i \alpha}.
\eea
This is nothing but the determinant of the twisted Wilson fermion matrix $\mathcal{W}_{\alpha}^F$
\beq
Z_\alpha^F = \det \left[\mathcal{W}_\alpha^F \right].
\eeq
For periodic boundary conditions ($\alpha = 0$ case) this is in agreement with the expression obtained in Ref. \cite{Catterall:2000rv}. 

The full partition function takes the form
\beq
\label{eqn:lat-pf-bos}
Z_\alpha = \left( \frac{1}{\sqrt{2\pi}} \right)^T \int \left(\prod_{k = 0}^{T-1} d\phi_k \right) ~ \exp \left[ - \mcS^{\rm ~eff}_\alpha \right],
\eeq
with
\bea
\label{eqn:lat-eff-action}
{\mcS_{\alpha}}^{\text{eff}} &=& {\mcS}^{B} - \ln \left( {\rm det} \left[ \mathcal{W}_\alpha^F \right] \right) \nn \\ 
&=& \sum_{k = 0}^{T-1} \hf \bigg( \phi_k - \phi_{k-1} + \Xi'_k  \bigg)^2 - \ln \left( \prod_{k = 0}^{T-1} \left(1 + \Xi^{''}_{kk} \right) - e^{i \alpha} \right).
\eea

Given an observable $\mathcal{O}$, we can compute its expectation value as
\bea
\label{eqn:lat-exp-obs}
\langle \mathcal{O} \rangle &=& \lim_{\alpha \to 0} \langle \mathcal{O} \rangle_\alpha \nn \\
&=& \lim_{\alpha \to 0} \frac{1}{Z_\alpha} \left( \frac{1}{\sqrt{2\pi}} \right)^T \int \left(\prod_{k = 0}^{T-1} d\phi_k \right) \mathcal{O} ~ \exp \left[{ - \mcS_{\alpha}^{\text{eff}}} \right].
\eea

Note that the gradient of the action has to be computed to update the field configurations in the CL method. The drift term, given by the negative of the gradient of the action, contains the fermion determinant in the denominator, whose zeroes in the complexified space cause the subtlety for the conditions required for the justification of CL method. The dynamical variables may come close to the singularity of the drift term (the singular drift problem). A recent study highlighted that such a problem is not restricted to logarithmic singularities but is rather generic and may arise where the stochastic process involves a singular drift term \cite{Nishimura:2015pba}.

\subsection{Correlation functions}
\label{corr-fn}

Using the expression given in Eq. \eqref{eqn:lat-exp-obs} we can compute the correlation functions. The bosonic and fermionic correlation functions are defined as
\beq
G^B_\alpha (k) \equiv \langle \phi_0 \phi_k \rangle_\alpha
\eeq
and
\beq
G^F_\alpha(k) \equiv \langle \psib_0 \psi_k \rangle_\alpha,
\eeq
respectively, at the site $k$.

The fermionic correlation function can be shown to be 
\beq
\langle \psib_0 \psi_k \rangle_\alpha = \frac{1}{Z_\alpha} \left( \frac{1}{\sqrt{2\pi}} \right)^T \int \left( \prod_{i = 0}^{T-1} d\phi_i \right) \underbrace{ \left( -\frac{{ \langle \psib_0 \psi_k \rangle^F} }{ \det \left[ \mathcal{W}_\alpha^F \right] } \right) }_{\left[ \psib_0 \psi_k \right]^L_\alpha} ~ \exp \left[ -{ S_\alpha }^{\text{eff}} \right].
\eeq
Upon comparison with Eq. \eqref{eqn:lat-exp-obs} we define $\left[ \psib_0 \psi_k \right]^L_\alpha$ as our Langevin observable corresponding to the fermionic correlator $\langle \psib_0 \psi_k \rangle_\alpha$. That is,
\beq
{\left[\psib_0 \psi_k \right]^L_\alpha} = \left(-  \frac{ \prod_{i = k + 1}^{T-1} \left[1 + \Xi^{''}_{ii} \right] }{ \prod_{i = 0}^{T-1} \left[ 1 + \Xi^{''}_{ii} \right] - e^{i \alpha}} \right).
\eeq

Now, for the bosonic correlation function, the computation is rather straightforward. The Langevin observable is the bosonic correlation function itself. For the $k$-th lattice site, ${ \left[ \phi_0 \phi_k \right]^L} = \phi_0 \phi_k $, such that
\bea
\langle \phi_0 \phi_k \rangle_\alpha = \frac{1}{ Z_\alpha } \left( \frac{1}{ \sqrt{2 \pi} } \right)^T \int \left( \prod_{i = 0}^{T-1} d\phi_i \right) \phi_0 \phi_k ~ \exp \left[ { - S_\alpha }^{ \text{eff}} \right]. 
\eea

\subsection{Ward identities}
\label{ward-id}

Another set of observables that would help us in the investigations on SUSY breaking is the Ward identities. For the supersymmetric variation of the fields, Eqs. \eqref{eqn:lat-Wilson-Q} and \eqref{eqn:lat-Wilson-Qbar}, the invariance of the lattice action guides us to a set of Ward identities that connect the bosonic and fermionic correlators. The partition function in Eq. \eqref{eqn:lat-pf}, upon addition of the source terms  ($J, \theta, \overline{\theta}$), becomes
\bea
\label{eqn:lat-pf-source}
Z_\alpha \left( J, \theta, \overline{\theta} \right) &=& \left( \frac{1}{ \sqrt{2 \pi} } \right)^T \int \left( \prod_{k = 0}^{T-1} d\phi_k d\psi_k d\psib_k \right) \nn \\ 
&&\hspace{2cm} \times \exp \left[{ - \mcS_\alpha + \sum_{k = 0}^{T-1} \left( J_k \phi_k + \theta_k \psib_k + \overline{\theta}_k \psi_k \right) } \right].
\eea

It is easy to see that the variation of the partition function under the $\Q$-transformations vanishes upon turning off the external sources. That is, $\Q Z_\alpha \left( J, \theta, \overline{\theta} \right) = 0$. In fact, the variation of any derivative of the partition function with respect to these external source terms also vanishes (upon turning off the sources). Taking the derivative of the partition function with respect to the source terms $J_j$ and $\theta_i $ we get following set of non-trivial supersymmetric Ward identities,
\beq
\langle \psib_i \psi_j \rangle + \langle N_i \phi_j \rangle = 0.
\eeq

We will consider
\beq
\label{eqn:lat-ward-iden}
\mathscr{W}_1 : 
\begin{array}{l}
	\langle \psib_0 \psi_k \rangle  +  \langle  N_0 \phi_k \rangle = 0
\end{array}
\eeq
to probe the SUSY breaking.

\section{Complex Langevin simulations}
\label{lattice-simulation}

Let us look at the relevant observables we used in the simulations. One crucial observable is the expectation value of the auxiliary field
\bea
\B_\alpha = -i \left( \nabla^S_{ij} \phi_j + \Omega'_i \right)  = -i \left( \nabla^{-}_{ij} \phi_j + \Xi'_i \right).
\eea

Studies have shown that the auxiliary field expectation value can be used as an order parameter to reliably predict dynamical SUSY breaking \cite{Kuroki:2009yg,Kuroki:2010au, Joseph:2019sof}. The non-vanishing (vanishing) nature of the auxiliary field indicates that SUSY is broken (preserved) in the system. That is,
\begin{equation}
\langle \mathcal{B} \rangle =\lim_{\alpha \to 0} \langle \mathcal{B}_{\alpha} \rangle  
\begin{cases}
\neq 0 & \text{SUSY broken} \\
=0 & \text{SUSY preserved}. 
\end{cases}
\end{equation}
However, this vanishing nature could be accidental for some models. In \cite{Kuroki:2009yg}, higher powers of $\B$ were considered to confirm SUSY breaking for zero-dimensional models. Thus, we also analyze other significant observables to confirm SUSY breaking predictions.

We next consider the bosonic action $\mcS^B_\alpha$. It has been studied that for exact lattice SUSY, the expectation value of the bosonic action is independent of the interaction couplings \cite{Catterall:2001fr}. Thus, the bosonic action expectation value simply counts the number of degrees of freedom on the lattice \cite{Catterall:2001fr,Catterall:2009it}. That is, $\langle \mathcal{S}^B \rangle = \hf N_{\rm d.o.f}$. In supersymmetric quantum mechanics, it is expected that, $\langle \mcS \rangle = T$, and $\langle \mcS^B \rangle = T/2$,  where $T$ is the number of sites. Thus we have
\beq
\langle \mcS^B \rangle =\lim_{\alpha \to 0} \langle \mathcal{S}^B_\alpha \rangle
\begin{cases}
	\neq {T}/{2} & \text{SUSY broken} \\
	=T/2 & \text{SUSY preserved}. 
\end{cases}
\eeq

The third indicator is the equality of the fermionic and bosonic mass gaps. The mass gaps can be extracted either by a $\cosh \big[ m a( t - \frac{T}{2}) \big]$ fit for the $t$-th lattice site, or a simple exponential fit over say, the first or last $T/4$ time slices of the respective correlation functions \cite{Catterall:2001fr, Giedt:2004vb}.

The last set of observables involve the Ward identity. They can be used to confirm exact lattice supersymmetry successfully. See Refs. \cite{Catterall:2000rv, Catterall:2001fr, Catterall:2003wd, Kadoh:2018ele}. We expect that $\mathscr{W}_1$, given in Eq. \eqref{eqn:lat-ward-iden}, to hold (not to hold) for theories with SUSY preserved (broken). That is,
\beq
\lim_{\alpha \to 0} \mathscr{W}_1:
\begin{cases}
	-\langle  \psib_0 \psi_k \rangle_\alpha  \neq \langle  N_0  \phi_k \rangle_\alpha & \text{SUSY broken} \\
	-\langle  \psib_0 \psi_k \rangle_\alpha = \langle  N_0  \phi_k \rangle_\alpha  &  \text{SUSY preserved}. 
\end{cases}
\eeq

Only in the limit $\alpha \to 0$ we can comment, using the above set of observables, if the system possesses exact lattice SUSY. Since the partition function is a well-defined quantity for models with SUSY preserved, as expected, we were able to compute the normalized expectation values of observables, and hence perform numerical investigations for $\alpha = 0$ (PBC) case. The issue in working without the twist field arises only in models where SUSY is spontaneously broken, since the partition function vanishes, and normalized expectation values of the observables are ill-defined. Hence in Sec. \ref{subsec:gen-sps} we perform CL simulations for various values of the twist parameter to verify the consistency of our results for the $\alpha = 0$ case. 

\subsection{Supersymmetric anharmonic oscillator}
\label{subsec:susy-anho}

The model we consider in this section, the supersymmetric anharmonic oscillator, has a real action. Our goal is to put the simulation code to test by comparing our results with those given in Ref. \cite{Catterall:2000rv}.

The model has the potential 
\beq
\label{eqn:sim-anho}
\Xi(\phi) = \hf m \phi^2 + \qtr g \phi^4.
\eeq
This model has been investigated in great detail with the help of Hybrid Monte Carlo (HMC) algorithm in the lattice and non-lattice formalisms, respectively, in Ref. \cite{Catterall:2000rv} and Ref. \cite{Hanada:2007ti}. It was concluded that SUSY is preserved in this model for a finite value of the coupling. In the case with $\alpha = 0$, the action is real and the evolution of field configurations in the system is governed by real Langevin dynamics. 

\begin{figure}[tbp]
	\begin{center}
	{

	\includegraphics[width=.55\textwidth,origin=c,angle=0]{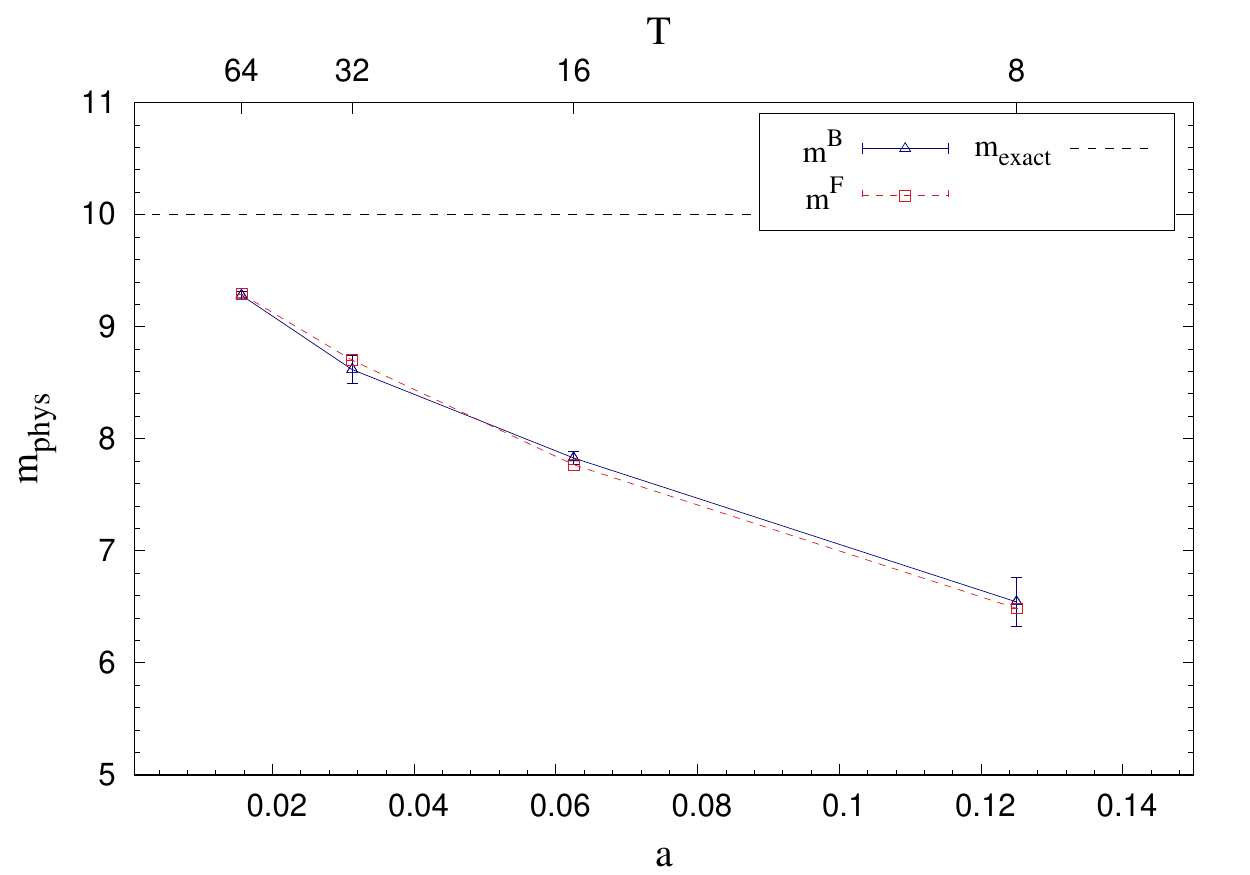}
	
	\caption {Bosonic and fermionic mass gaps for SUSY harmonic oscillator with physical parameters $m_{\rm phys} = 10$ and $g_{\rm phys} = 0$ versus lattice spacing ($a$) and lattice size ($T$).}
	\label{fig:lat-susy-ho-m10g0-massgaps}	
}
\end{center}
\end{figure}

First, we simulate SUSY harmonic oscillator for physical parameters $m_{\rm phys} = 10$ and $g_{\rm phys} = 0$, and $\alpha = 0$. Simulations were performed for different lattice spacings keeping the (physical) circle size $\beta = 1$. Figure \ref{fig:lat-susy-ho-m10g0-massgaps} shows bosonic (blue triangle) and fermionic (red square) physical mass gaps versus lattice spacing ($a$) and lattice size ($T$). Black dashed line shows the continuum value of SUSY harmonic oscillator mass gaps for the physical parameters $m_{\rm phys} = 10$, $g_{\rm phys} = 0$, that is, $m_{\rm exact} = 10$. We see that boson and fermion masses are degenerate within statistical errors, and furthermore, as lattice spacing $a \to 0$, the common mass gap approaches the correct continuum value. The simulations confirm that the free action has an exact SUSY at finite lattice spacing, which is responsible for the degenerate mass gaps. 

Now we simulate the SUSY anharmonic oscillator for physical parameters $m_{\rm phys} = 10$ and $g_{\rm phys} = 100$, and $\alpha = 0$. Simulations were performed for different lattice spacings keeping the (physical) circle size $\beta = 1$. In table \ref{tab:lat-susy-anho-m10g100} we provide the bosonic and fermionic mass gaps. Here also we have $m^B_{\text{phys}} \approx m^F_{\text{phys}}$ indicating that SUSY is preserved in the model. Table \ref{tab:lat-susy-anho-m10g100-action-B} contains the expectation values of the auxiliary field $\B_{\alpha}$ and bosonic action $\mcS^B_{\alpha}$. The mean expectation value $\langle \B \rangle $ vanishes in the simulations and thus indicate exact lattice SUSY. This table also contains the expectation value of the bosonic action $\mcS^B_{\alpha}$. For this model, we observe $\langle \mcS^B \rangle = \hf T$, and it was independent of physical parameters $g_{\rm phys}$ and $m_{\rm phys}$, which again suggests SUSY is preserved in this model. 

\begin{figure}[tbp]
	\centering
	\includegraphics[width=.49\textwidth,origin=c,angle=0]{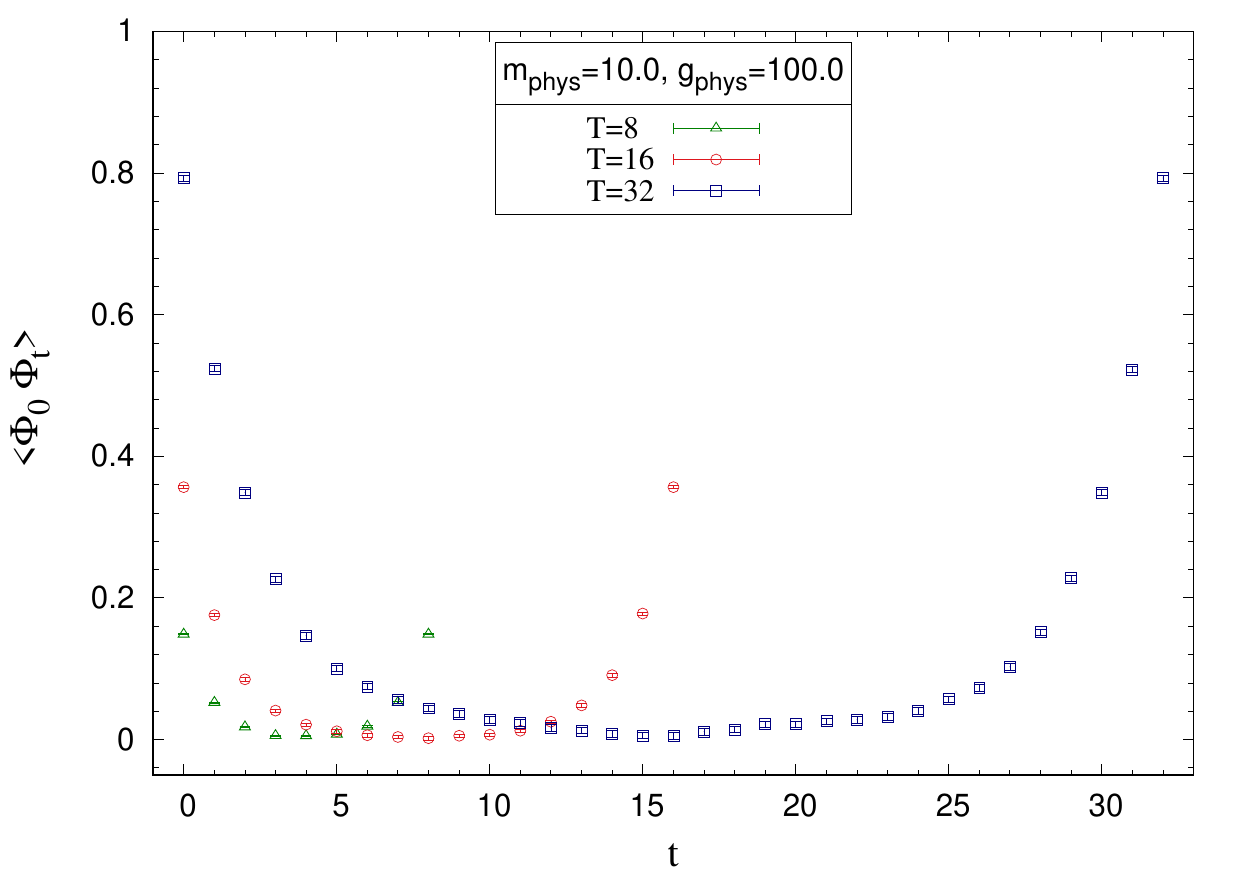}
	\includegraphics[width=.49\textwidth,origin=c,angle=0]{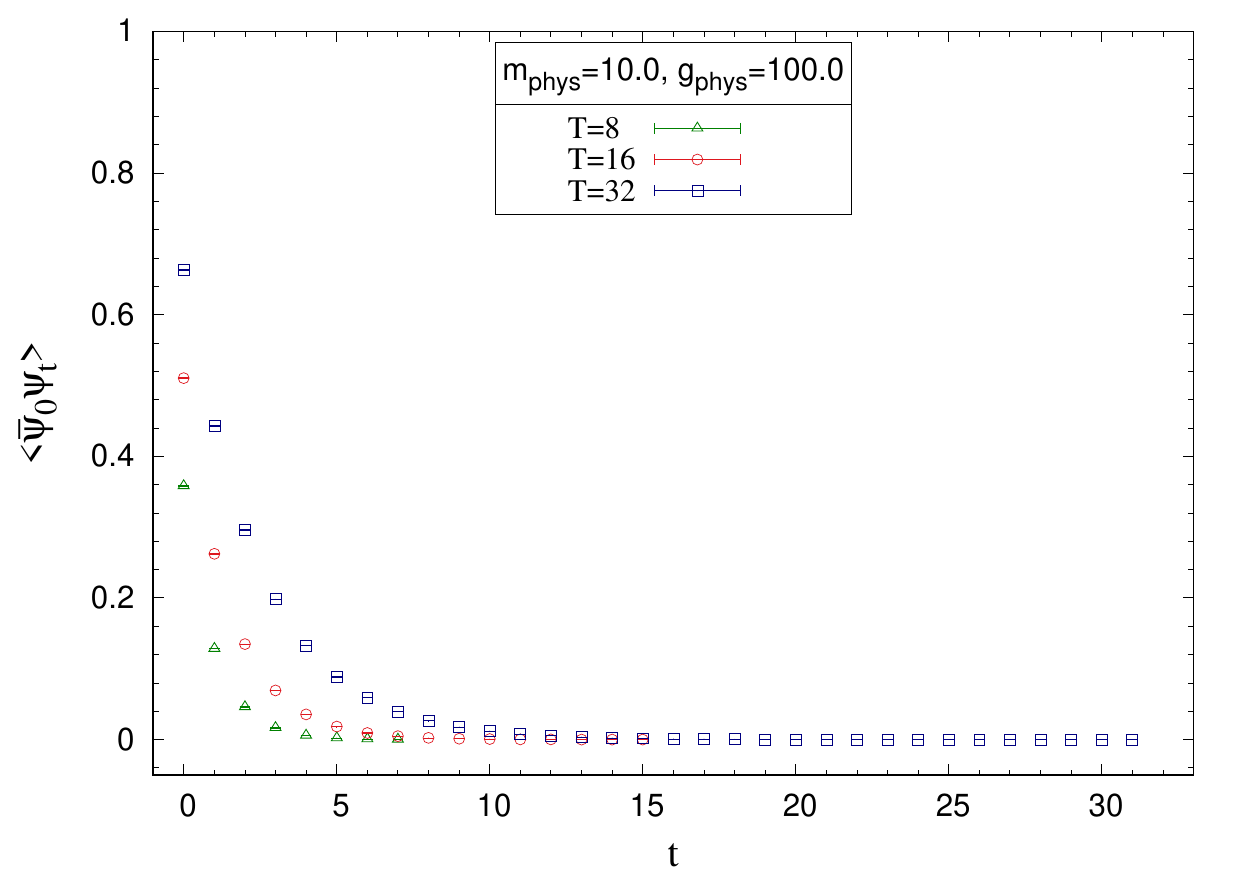}
	
	\caption{Bosonic (left) and fermionic (right) correlation functions for SUSY anharmonic oscillator. The parameters used are $m_{\rm phys} = 10$ and $g_{\rm phys} = 100$.}
	\label{fig:lat-susy-anho-m10g100-corr}	
\end{figure}

\begin{figure*}[tbp]
	\centering
	\includegraphics[width=.55\textwidth,origin=c,angle=0]{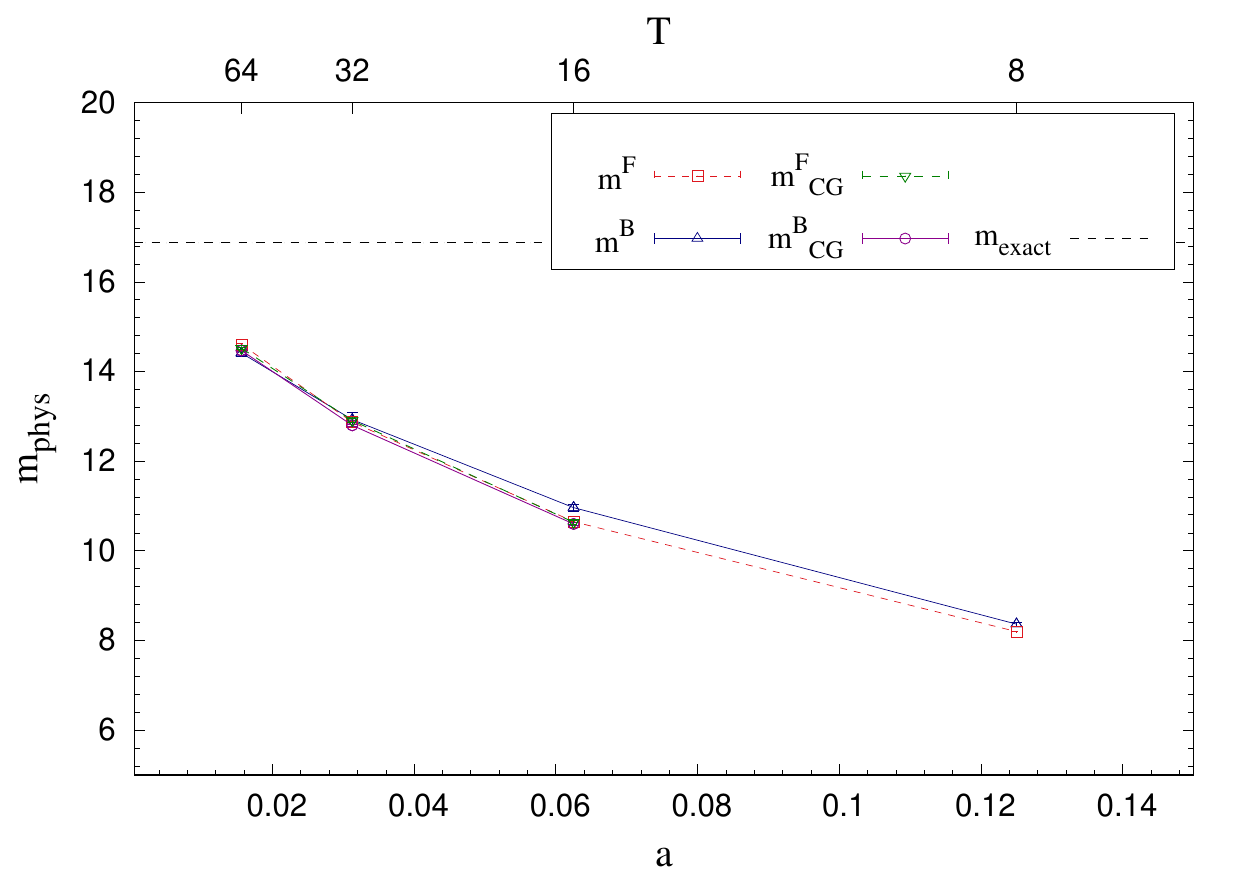}
	
	\caption {Bosonic and fermionic mass gaps for SUSY anharmonic oscillator. The parameters used are $m_{\rm phys} = 10$ and $g_{\rm phys} = 100$. Here $m^{B}_{\rm CG}$ and $m^{F}_{\rm CG}$, respectively represent bosonic and fermionic mass-gap results from Ref. \cite{Catterall:2000rv}.}
	\label{fig:lat-susy-anho-m10g100-massgaps}	
\end{figure*}

\begin{figure*}[tbp]
	\centering
	\includegraphics[width=.49\textwidth,origin=c,angle=0]{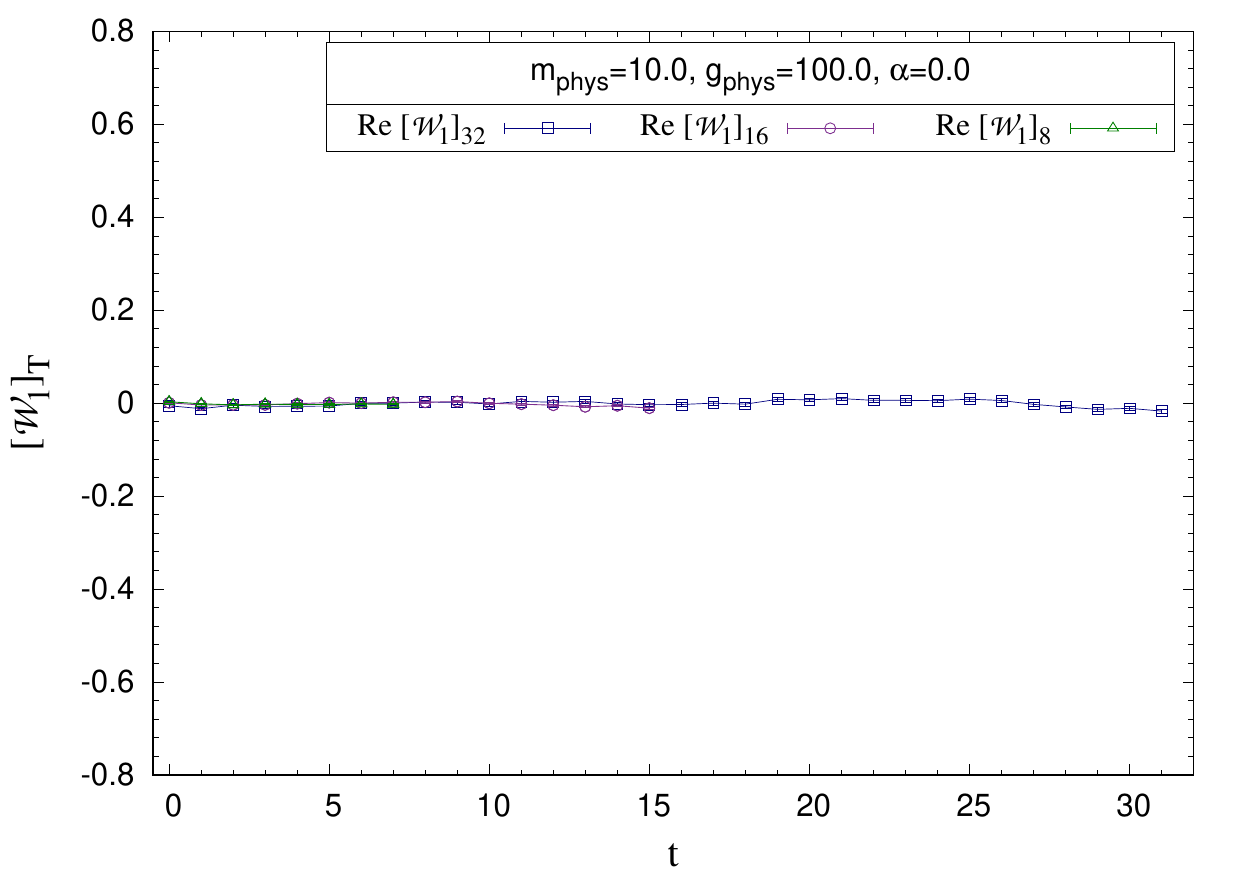}
	\includegraphics[width=.49\textwidth,origin=c,angle=0]{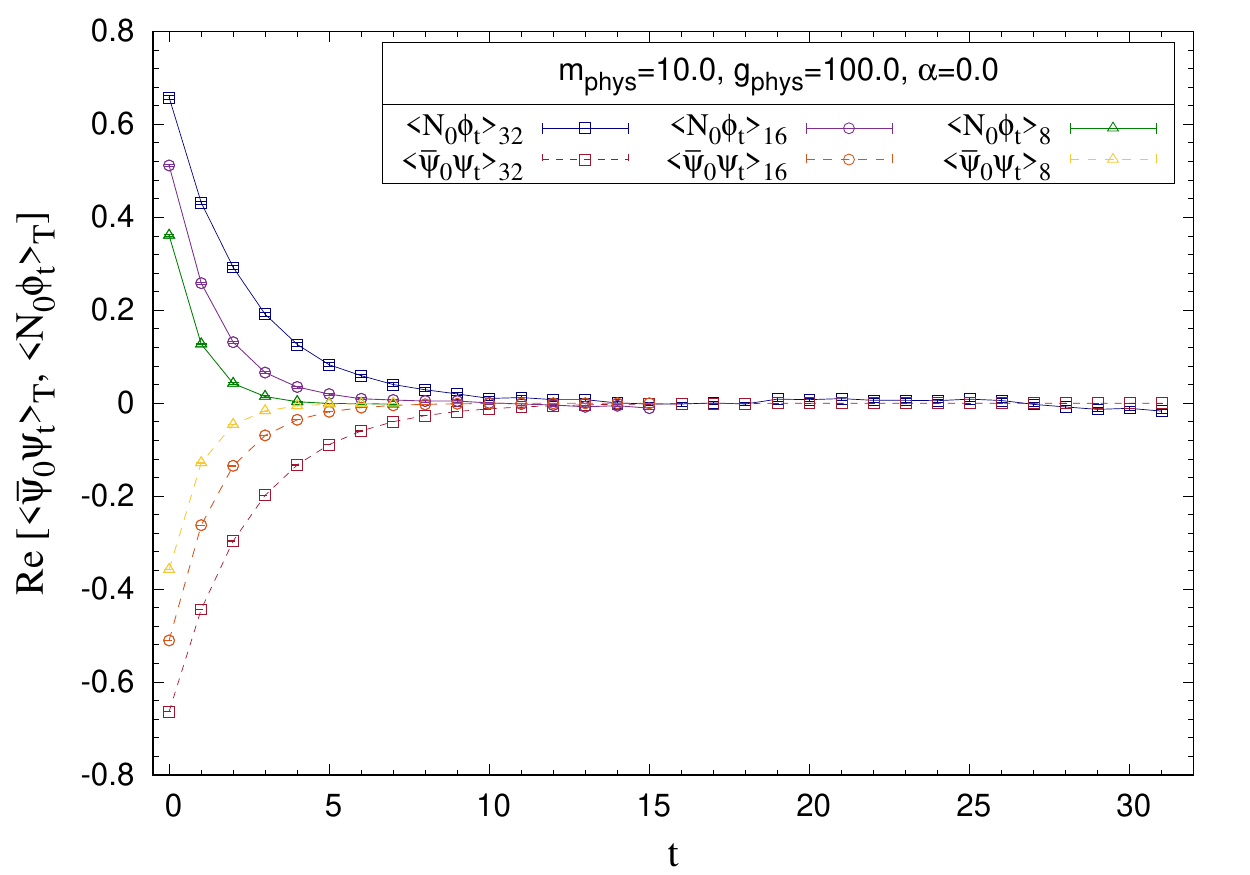}
	
	\caption {Real part of the Ward identity (left) and real part of the bosonic and fermionic contributions to the Ward identity (right), versus lattice site $t$ for lattices with $T$. Simulations were performed for SUSY anharmonic oscillator with physical parameters $m_{\rm phys} = 10$ and $g_{\rm phys} = 100$ for lattice sizes $T = 8, 16$, and $32$.}
	\label{fig:lat-susy-anho-m10g100-ward}	
\end{figure*}

Figure \ref{fig:lat-susy-anho-m10g100-corr} shows the bosonic (left) and fermionic (right) correlation functions (used to compute the respective mass gaps) versus lattice site ($t$) for various lattice size ($T$) for the SUSY anharmonic oscillator. In Fig. \ref{fig:lat-susy-anho-m10g100-massgaps} we show the bosonic and fermionic physical mass gaps versus lattice spacing ($a$) and lattice size ($T$). Here we also compare our results with those obtained by Catterall and Gregory \cite{Catterall:2000rv}, ($m^{B}_{\rm CG}$, $m^{F}_{\rm CG}$), and find excellent agreement. Black dashed line shows the continuum value of SUSY anharmonic oscillator mass gaps for the physical parameters $m_{\rm phys} = 10$, $g_{\rm phys} = 100$ that is $m_{\rm exact}$ = 16.865 \cite{Bergner:2007pu}. We see that boson and fermion masses are degenerate within statistical errors, and furthermore, as lattice spacing $a \to 0$, the common mass gap approaches the correct continuum value. In Fig. \ref{fig:lat-susy-anho-m10g100-ward} we plot the real part of Ward identity $\mathscr{W}_1$  (left), and its bosonic and fermionic contributions (right), given in Eq. \eqref{eqn:lat-ward-iden}, versus the lattice site $t$ for lattices with $T$ values. We observe that the respective bosonic and fermionic contributions cancel each other out within statistical errors, and hence $\mathscr{W}_1$ is satisfied. Our results confirm that the SUSY anharmonic oscillator has an exact SUSY, which is responsible for the degenerate mass gaps. 

\begin{table}[tbp]
	\begin{center}
	{\small	
	\begin{tabular}{|l	r|	l	r	|l	r|} 
		\hline
		$T$ &  $a = T^{-1}$  &  $m^B$   &  $m^B_{\text{phys}} = a^{-1} m^B$&  $m^F$   &  $m^F_{\text{phys}} = a^{-1} m^F$  \\	
		\hline
		\hline
		8   & 0.125   	 & $1.0457(65)$	  &$8.3656(520)$	&   $1.0247(4)$	  &$8.1976(32)$    	\\
		16  & 0.0625   	 & $0.6852(45)$  &  $10.9632(720)$	& $0.6657(1)$	  &$10.6512(16)$	   \\
		32  & 0.03125  	 & $0.4040(54)$  & $12.9280(1664)$	 & $0.4023(2)$	  &$12.8736(64)$	\\
		64  & 0.015625 	 & $0.2252(13)$  &  $14.4128(832)$	& $0.2282(3)$	  &$14.6048(192)$	\\
		\hline
	\end{tabular}}
	\caption{Bosonic and fermionic mass gaps for SUSY anharmonic oscillator. The parameters used are $m_{\rm phys} = 10$ and $g_{\rm phys} = 100$.}
	\label{tab:lat-susy-anho-m10g100}
	\end{center}
\end{table}

\begin{table}[tbp]
	\begin{center}
	{\small
	\begin{tabular}{|c |l	r|	c|	l	r|} 
		\hline
		$\Xi'(\phi)$	&$T$ &  $a = T^{-1}$  &  \makecell{ $\alpha$ } &  $~\langle \B_{\alpha} \rangle$   &  $~\langle \mathcal{S}^B_{\alpha} \rangle$ \\   
		\hline
		\hline
		$m \phi + g \phi^3$
		&$8$ &  $0.1250$  & $0.00$ &  $0.0(0) -i0.0008(38)$ & $4.0672(67) + i 0.0(0)$ \\
		&$16$ &  $0.0625$  & $0.00$ &  $0.0(0) +i0.0003(68)$ & $8.0698(95) + i 0.0(0)$ \\
		&$32$ &  $0.03125$  & $0.00$ &  $0.0(0) -i0.0038(131)$ & $16.1589(147) + i 0.0(0)$ \\
		&$64$ &  $0.015625$  & $0.00$ &  $0.0(0) -i0.0162(245)$ & $32.2293(252) + i 0.0(0)$ \\
		\hline
	\end{tabular}}
	\caption{Expectation value of the auxiliary field $\B_\alpha$ and the bosonic action $\mcS^B_\alpha$ for SUSY anharmonic oscillator. The parameters used are $m_{\rm phys} = 10$ and $g_{\rm phys} = 100$.}
	\label{tab:lat-susy-anho-m10g100-action-B}
		\end{center}
\end{table}

\subsection{General polynomial potential}
\label{subsec:gen-sps}

As a check of our code we consider the model with a degree-$k$ polynomial potential with real coefficients
\beq
{\Xi{'}}^{(k)} = g_k \phi^k + g_{k-1} \phi^{k-1} + \cdots + g_0.
\eeq
In these systems, it is well known that SUSY is spontaneously broken if the count of zeroes of the potential is even, that is when ${\Xi}^{(k)}(-\infty)$ and ${\Xi}^{(k)}(+\infty)$ have opposite signs \cite{Witten:1981nf}. For simplicity, we assume the form $g_k = g$, $g_{k-1} = \cdots = g_2 = 0$, $g_1 = m$ and $g_0 = g \mu^2$. Then, for degree $k = 4$ and $5$, we have
\bea
\label{eqn:lat-gen-pot-k4}
{\Xi{'}}^{(4)} &=& g \phi^4 + m \phi + g \mu^2, \\
\label{eqn:lat-gen-pot-k5}
{\Xi{'}}^{(5)} &=& g \phi^5 + m \phi + g \mu^2.
\eea

Using complex CL method, we confirm the analytical prediction that SUSY is broken for even-($k = 4$) and preserved for odd-($k = 5$) degree real-polynomial potentials.

For even-degree potential we encounter {\it singular drift problem} when $\alpha = 0$. This shows up in the simulations as the complex Langevin correctness criteria involving the probability of absolute drift not being satisfied. For $\alpha = 0$, we obtained a power law decay instead of an exponential decay. Therefore we have performed simulations for various non-zero $\alpha$. Our simulations suggest that above a particular $\alpha$ value the correctness criteria is satisfied and the probability of absolute drift falls off exponentially. 

\begin{figure*}[tbp]
	\centering
	\includegraphics[width=.45\textwidth,origin=c,angle=0]{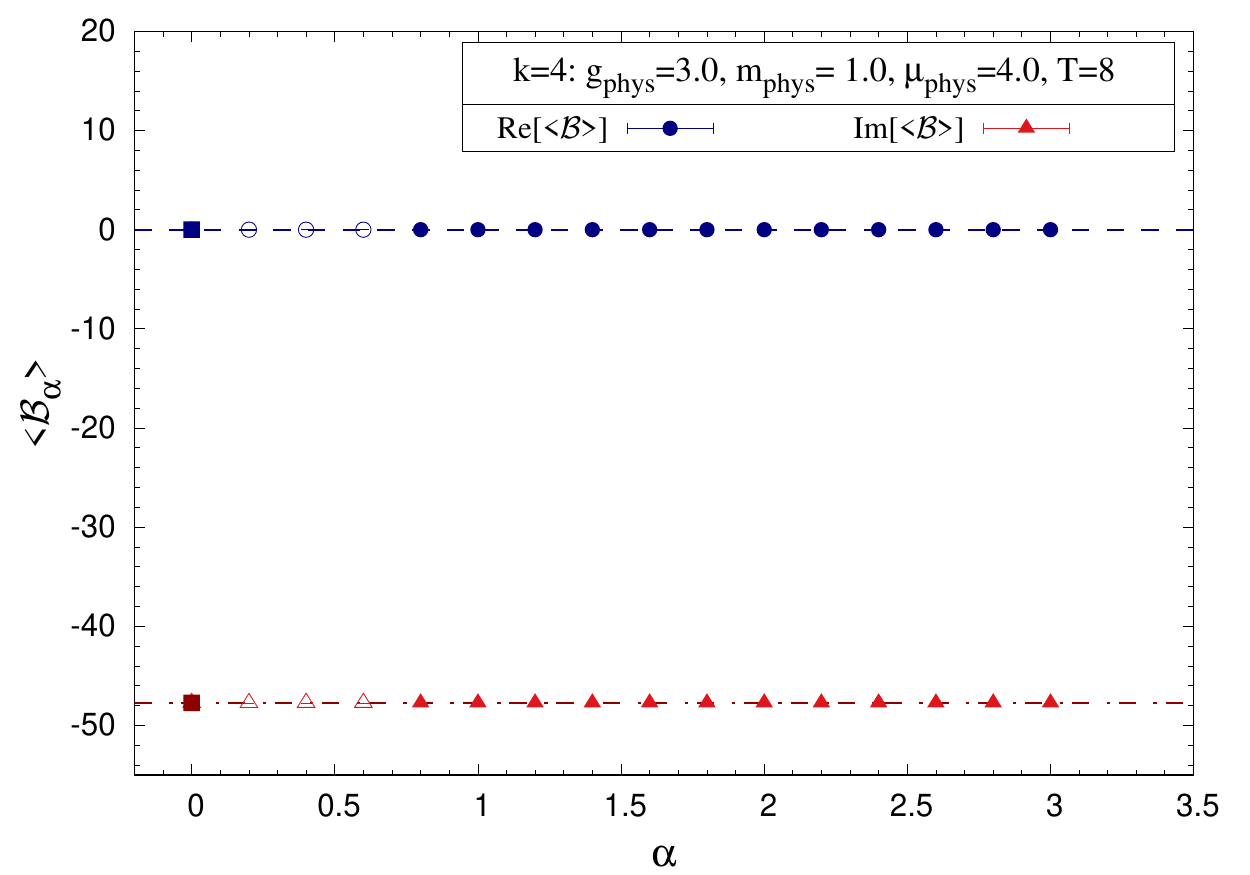} $~~~$ \includegraphics[width=.45\textwidth,origin=c,angle=0]{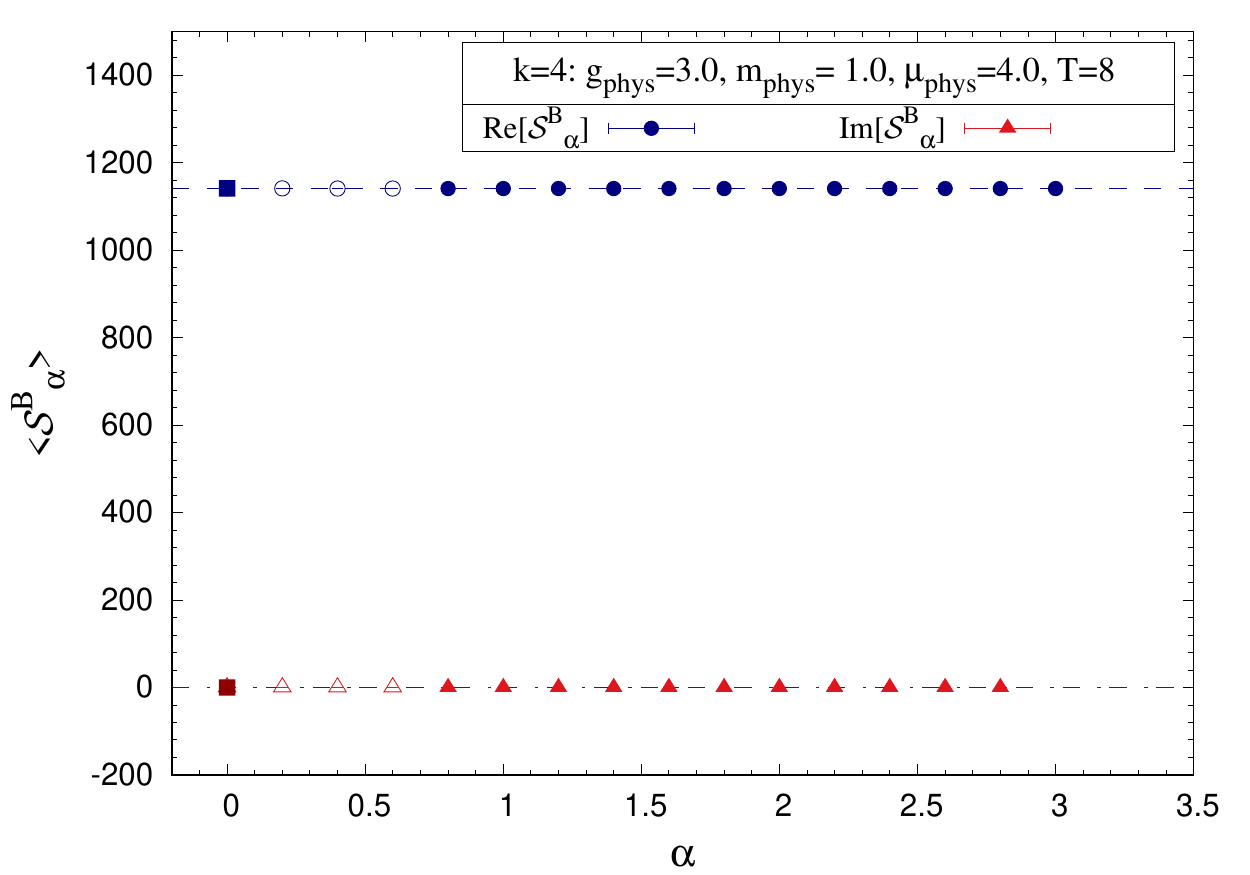}
	
	\caption{ Expectation values of auxiliary field $\B_\alpha$ (left) and the bosonic action $\mathcal{S}^B_\alpha$ (right) for various $\alpha$ values on a lattice with $T= 8$. The simulations are for the model with even-degree real polynomial potential ${\Xi{'}}^{(4)}$ given in Eq. \eqref{eqn:lat-gen-pot-k4}. The parameters used are $g_{\rm phys} = 3$, $m_{\rm phys} = 1$, and $\mu_{\rm phys} = 4$. 
	}
	\label{fig:lat-susy-k4-b-sb}	
\end{figure*}

In Fig. \ref{fig:lat-susy-k4-b-sb} we show the expectation value of the auxiliary field $\langle \B_\alpha \rangle$ (left) and the bosonic action $\langle \mcS^{B}_\alpha \rangle$ (right) on a $T = 8$ lattice for various $\alpha$ values. We consider the parameter space where CL simulations are justified and take the limit $\alpha \to 0$. The filled data points (red triangles for imaginary part and blue circles for real part) represent expectation values of observables for parameter space where CL can be trusted, while for unfilled data points CL correctness criteria is not satisfied. The lines represent a linear fit of observables for parameter space where CL simulations are justified and solid squares represent values of respective observables in the limit $\alpha \to 0$. These results indicate that $\langle \B  \rangle$ does not vanish. The expectation value of the bosonic action $\langle \mcS^B \rangle \neq \hf T$, and it is found to be dependent on the physical parameters. These results indicate that SUSY is broken for the even-degree real-polynomial superpotential. 

\begin{figure*}[tbp]
	\centering
	
	\includegraphics[width=.49\textwidth,origin=c,angle=0]{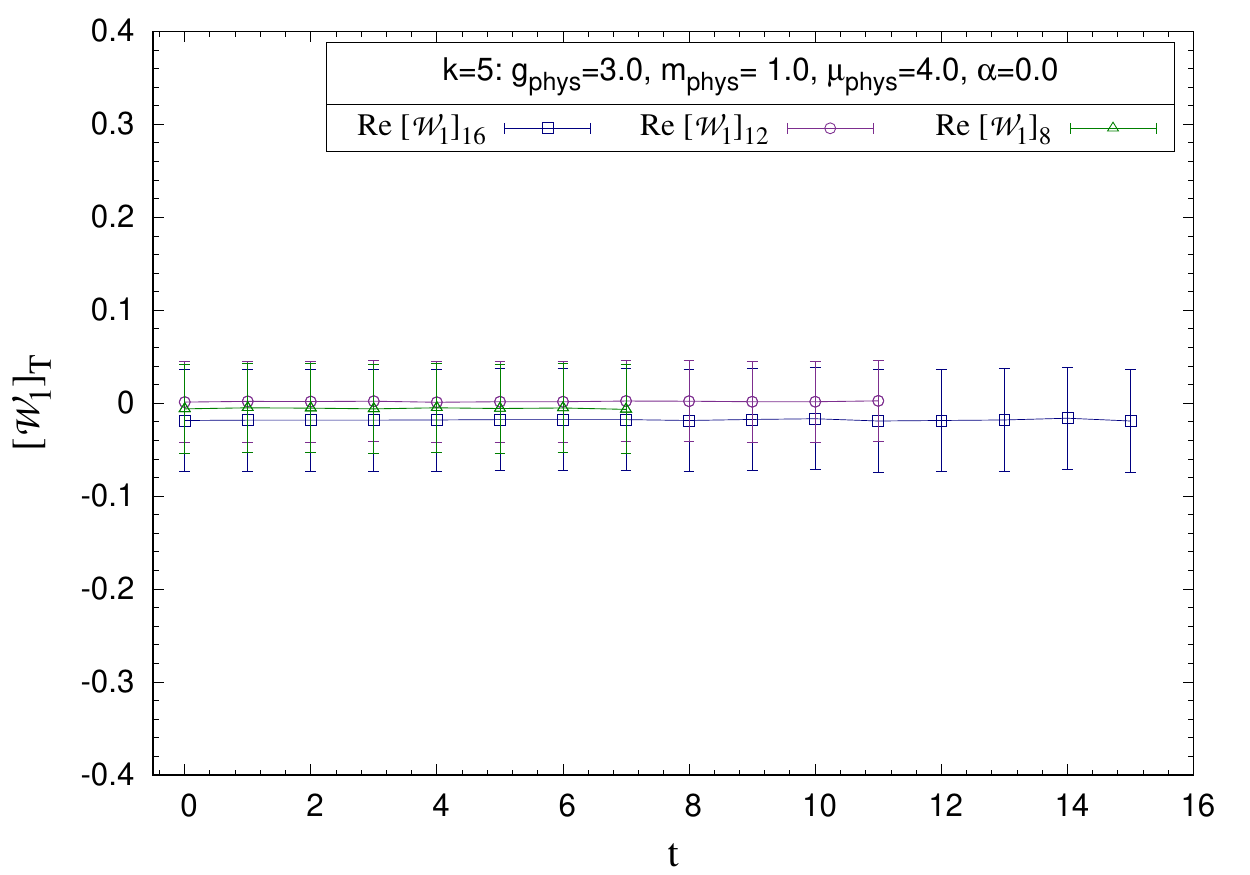}
	\includegraphics[width=.49\textwidth,origin=c,angle=0]{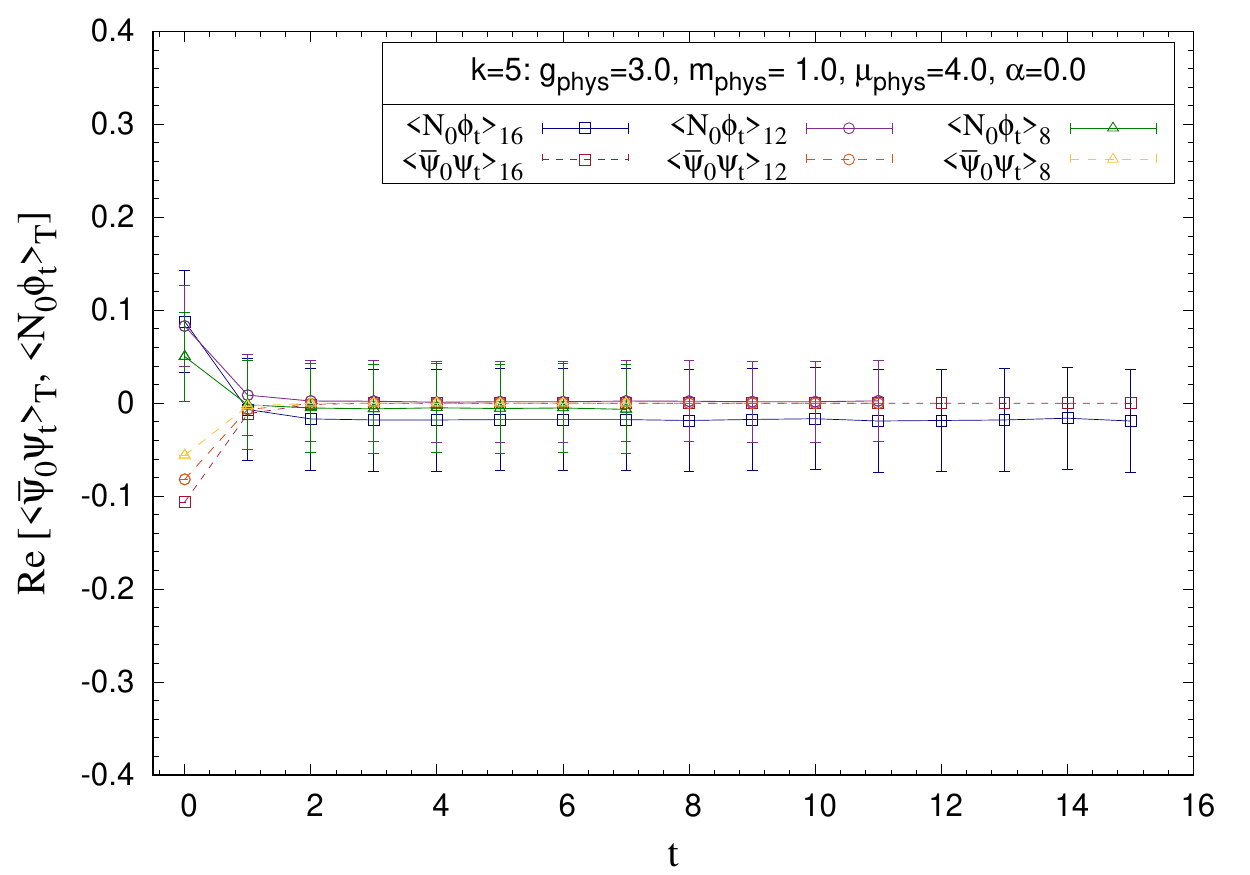}
	
	\caption {Real part of the Ward identity (left) and real part of bosonic and fermionic contributions to the Ward identity (right) for the model with odd-degree real polynomial potential. The parameters used are $g_{\rm phys} = 3$, $m_{\rm phys} = 1$, and $\mu_{\rm phys} = 4.0$ on lattices with $T = 8, 12$, and $16$.}
	\label{fig:k5ward}	
\end{figure*}

For the model with odd-degree potential, we observe that for $\alpha = 0$ the CL correctness criteria are satisfied. In table \ref{tab:k5bSb} we provide the simulation results for physical parameter $g_{\rm phys} = 3,~m_{\rm phys} = 1~ {\rm and}~ \mu_{\rm phys} = 4.0$, on lattices with $T = 8, 12,16$, and $\alpha = 0$. Our simulations gave vanishing value for the auxiliary field expectation value $\langle \B \rangle$. We also observed that the expectation value of bosonic action is $\langle \mcS^B \rangle = \hf T$ within errors. It is also independent of the coupling $g$. These results indicate that SUSY is preserved for these models. In Fig. \ref{fig:k5ward} we plot the Ward identities for this model. In these plots, on the left, we show the complete Ward identity (real part), and on the right, we present the respective bosonic and fermionic contributions to the Ward identity. For this model, the bosonic and fermionic contributions cancel each other out within statistical errors, and the Ward identity is thus satisfied. These results indicate unbroken SUSY in this model.

\begin{table}[tbp]
	\begin{center}
		{\small
			\begin{tabular}{|c| l r| c c|} 
				\hline
				$\Xi'(\phi)$&$T$ &  $a = T^{-1}$ &    $~\langle \B_{\alpha} \rangle$   &  $~\langle \mcS^B_{\alpha} \rangle$ \\   
				\hline
				\hline
				&${8}$	&${0.25}$	&
				$0.0000(0) 	+ i0.0042(101)$ & $4.0214(202)+ i0.0000(0)$ \\	
				${\Xi{'}}^{(5)}$
				&${12}$ &${0.125}$	
				& $0.0000(0) 	+ i0.0044(74)$ & $5.9942(176)+ i0.0000(0)$ \\ 	
				&${16}$	&${0.0833}$ &
				$0.0000(0) 	- i0.0026(86)$ & $8.0398(210)+ i0.0000(0)$\\
				\hline
			\end{tabular}
		}
		\caption{Expectation value of the auxiliary field $\B_{\alpha}$ and the bosonic action $\mcS^B_{\alpha}$ for the odd-degree superpotential ${\Xi{'}}^{(5)}$. The parameters used are $g_{\rm phys} = 3$, $m_{\rm phys} = 1$, and $\mu_{\rm phys} = 4.0$.}
		\label{tab:k5bSb}
	\end{center}
\end{table}

\subsection{$\mathcal{PT}$-symmetric models}
\label{subsec:pt-sps}

In this section, we simulate an interesting class of complex actions that exhibit, in addition to supersymmetry, $\mathcal{PT}$-symmetry. 

Quantum mechanics and quantum field theory are conventionally formulated using Hermitian Hamiltonians and Lagrangians, respectively. In recent years there has been an increasing interest in extensions to non-Hermitian quantum theories \cite{Bender:2002vv}, particularly those with $\mathcal{PT}$-symmetry \cite{Bender:1998ke, Bender:2005tb}, which have real spectra. Such theories have found applications in many areas such as optonics \cite{PhysRevLett.105.013903, Longhi_2017} and phase transitions \cite{Ashida_2017, Matsumoto_2020}. Recently, it has been shown that it is possible to carry over the familiar concepts from Hermitian quantum field theory, such as the spontaneous symmetry breaking and the Higgs mechanism in gauge theories, to $\mathcal{PT}$-symmetric non-Hermitian theories \cite{PhysRevD.98.045001, PhysRevD.99.045006, FRING2020114834}. In Ref. \cite{Alexandre:2020wki}, the authors constructed $\mathcal{PT}$-symmetric $\cN = 1$ supersymmetric quantum field theories in $3+1$ dimensions. There, they found that even though the construction of the models are explicitly supersymmetric, they offer a novel non-Hermitian mechanism for soft SUSY breaking.

Imposing $\mathcal{PT}$-symmetric boundary conditions on the functional-integral representation of the four-dimensional $- \lambda \phi^4$ theory can give a spectrum that is bounded below \cite{Bender:1999ek}. Such an interaction leads to a quantum field theory that is perturbatively renormalizable and asymptotically free, with a real and bounded spectrum. These properties suggest that a $- \lambda \phi^4$ quantum field theory might be useful in describing the Higgs sector of the Standard Model. 

We hope that our investigations would serve as a starting point for exploring the nonperturbative structure of these types of theories in higher dimensions.

The models we consider here have the following potential 
\beq
\label{eqn:lat-pt-symm-pot}
\Xi(\phi) = - \frac{g}{(2 + \delta)} \left(i \phi \right)^{\left(2 + \delta \right)},
\eeq
with $\Xi^{'}(\phi) = - i g ~ (i \phi)^{(1 + \delta)}$ and $\delta$ is a continuous parameter. The supersymmetric Lagrangian for this $\mathcal{PT}$-symmetric theory breaks parity symmetry, and it would be interesting to ask whether the breaking of parity symmetry induces a breaking of supersymmetry. This question was answered for the case of a two-dimensional model in Ref. \cite{Bender:1997ps}. There, through a perturbative expansion in $\delta$, the authors found that supersymmetry remains unbroken in this model. We investigate the absence or presence of non-perturbative SUSY breaking in the one-dimensional cousins of these models using CL method. Clearly, such an investigation based on path integral Monte Carlo fails since the action of this model can be complex, in general\footnote{In Ref. \cite{Dhindsa:2020ovr} it was shown that $\mathcal{PT}$ symmetry is preserved in supersymmetric quantum mechanics models with $\delta = 0, 2, 4$ using Monte Carlo simulations. See Refs. \cite{Kadoh:2015zza, Kadoh:2018ivg, Kadoh:2018ele, Kadoh:2019bir} for other related work on supersymmetric quantum mechanics on the lattice.}.

\subsection*{Even $\delta$ case}

We note that when $\delta = 0$ the model becomes the supersymmetric harmonic oscillator discussed in Sec. \ref{subsec:susy-anho}. 

In table \ref{tab:d2-4bSb} we provide the simulation results for $\delta = 2,~4$. Simulations were performed for physical parameter $g_{\rm phys} = 0.5$, on lattices with $T = 4, 8, 12$, and $\alpha = 0$. We noticed that the auxiliary field expectation value $\langle \B \rangle$ vanishes. We also observe that the expectation value of bosonic action is $\langle \mcS^B \rangle = \hf T$ within errors. It is also independent of the coupling $g$. These results indicate that SUSY is preserved in these models.

\begin{table}[tbp]
	\begin{center}
		{\small
			\begin{tabular}{|c| l r| c c|} 
				\hline
				$\Xi'(\phi)$&$T$ &  $a = T^{-1}$ &    $~\langle \B_{\alpha} \rangle$   &  $~\langle \mcS^B_{\alpha} \rangle$ \\   
				\hline
				\hline
				&${4}$	&${0.25}$	&
				$0.0000(0) 	+ i0.0005(282)$ & $2.0130(102)+ i0.0000(0)$ \\	
				$\delta = 2$
				&${8}$ &${0.125}$	
				& $0.0000(0) 	+ i0.0128(750)$ & $4.0326(157)+ i0.0000(0)$ \\ 	
				&${12}$	&${0.0833}$ &
				$0.0000(0) 	- i0.0071(263)$ & $6.0354(58)+ i0.0000(0)$\\
				\hline
				&${4}$	&${0.25}$	&
				$0.0000(0) 	+ i0.0167(679)$ & $1.9975(47)+ i0.0000(0)$ \\		
				$\delta = 4$
				&${8}$ &${0.125}$	
				& $0.0000(0) 	+ i0.0142(567)$ & $4.0058(54)+ i0.0000(0)$ \\	
				&${12}$	&${0.0833}$ &
				$0.0000(0) 	- i0.0309(1022)$ & $6.0018(74)+ i0.0000(0)$ \\ 
				\hline
			\end{tabular}
		}
		\caption{Expectation values of the auxiliary field $\B_\alpha$ and the bosonic action $\mcS^B_{\alpha}$ for the $\mathcal{PT}$-symmetric models with $\delta = 2, 4$.}
		\label{tab:d2-4bSb}
	\end{center}
\end{table}

\begin{figure*}[tbp]
	\centering	
	\includegraphics[width=.32\textwidth,origin=c,angle=0]{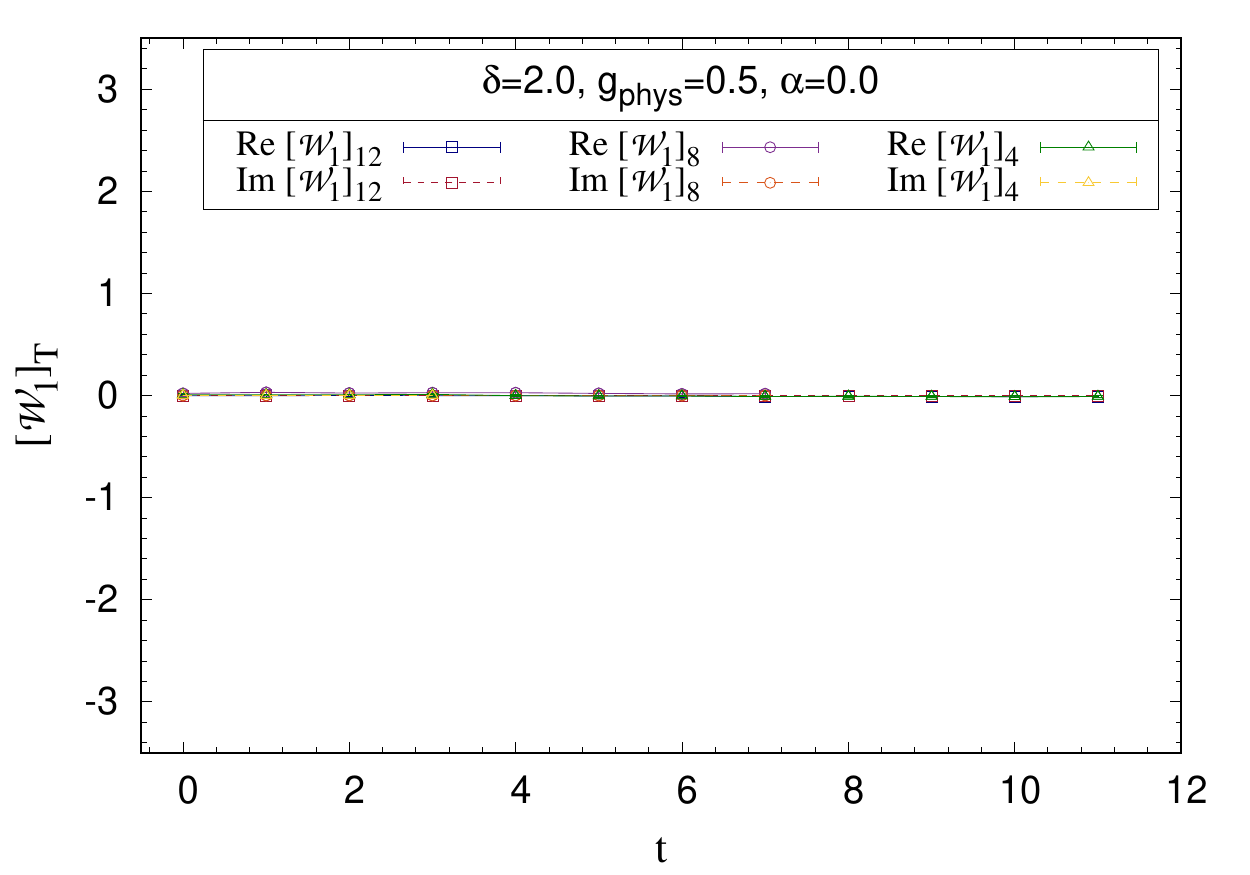}
	\includegraphics[width=.32\textwidth,origin=c,angle=0]{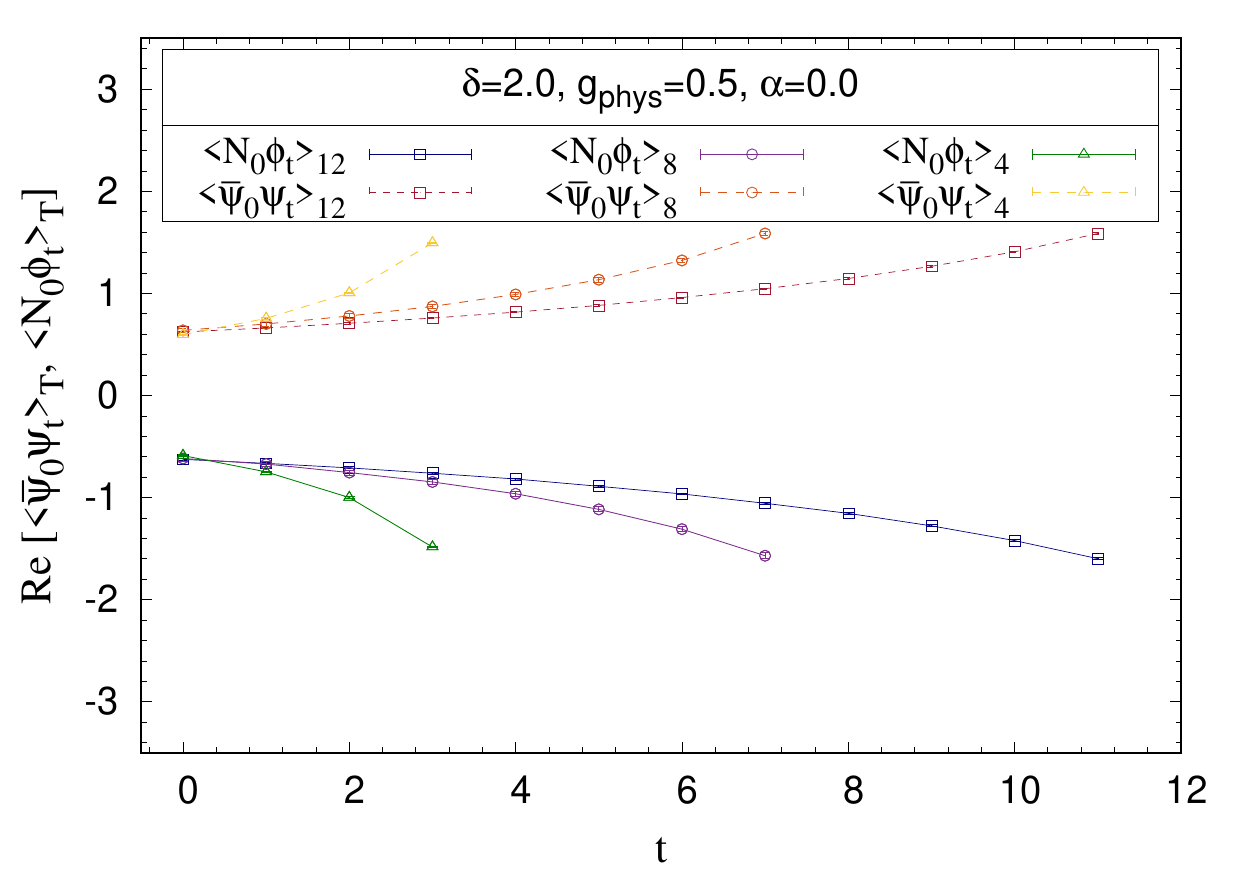}
	\includegraphics[width=.32\textwidth,origin=c,angle=0]{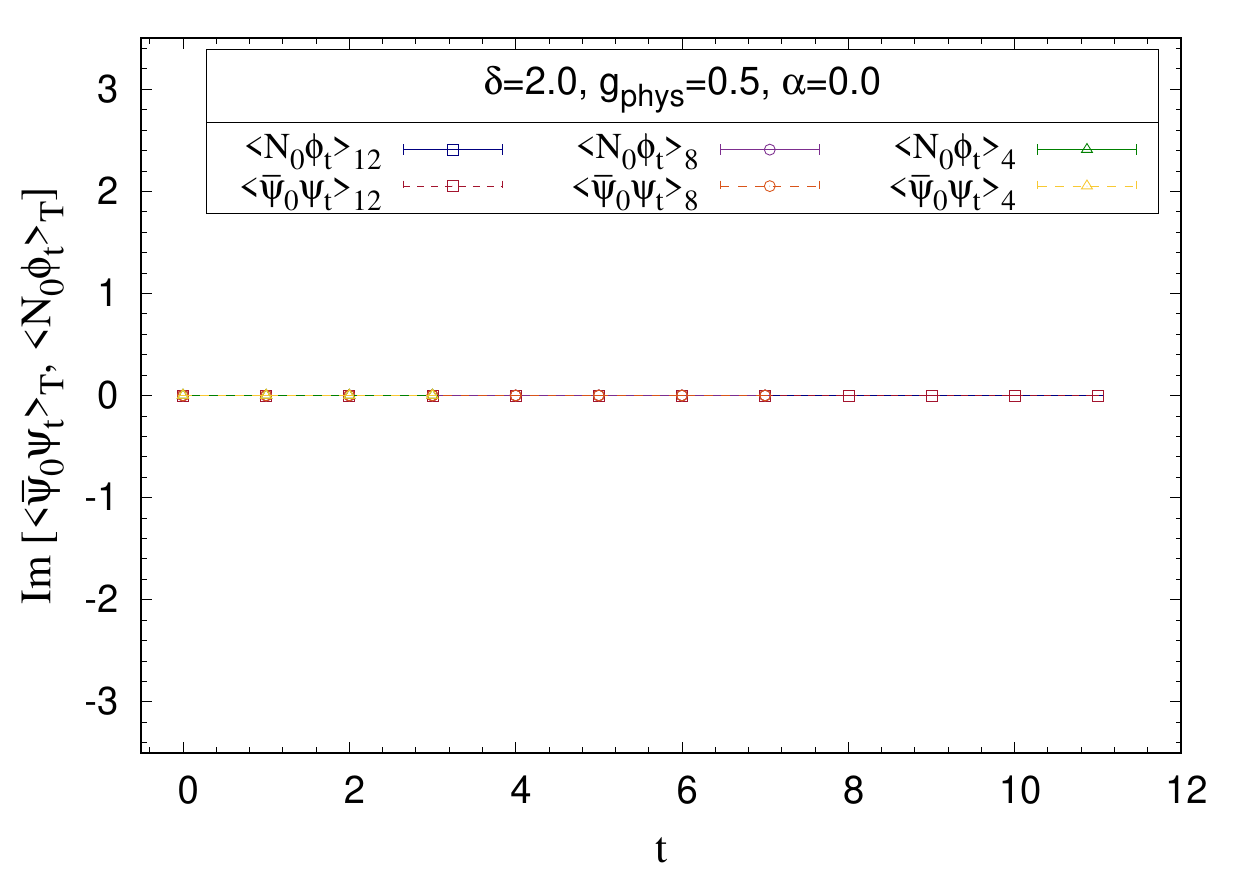}
	
	\caption {The  $\mathcal{PT}$-symmetric model with $\delta = 2$. The full Ward identity (left) and real (middle) and imaginary (right) parts of bosonic and fermionic contributions to Ward identity, for lattices with $T = 4, 8$, and $12$.}
	\label{fig:lat-susy-d2-T8-B-Sb}	
\end{figure*}

\begin{figure*}[tbp]
	\centering	
	\includegraphics[width=.32\textwidth,origin=c,angle=0]{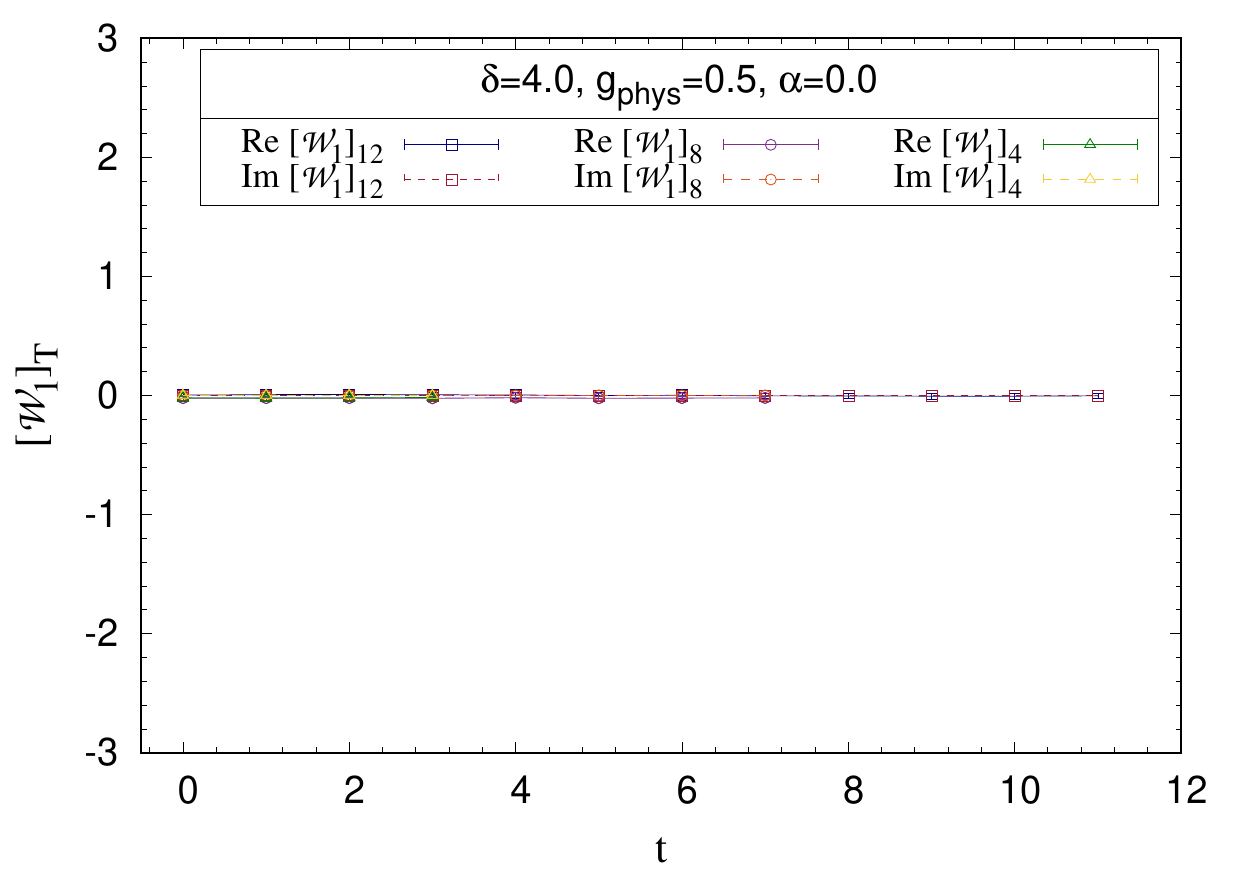}
	\includegraphics[width=.32\textwidth,origin=c,angle=0]{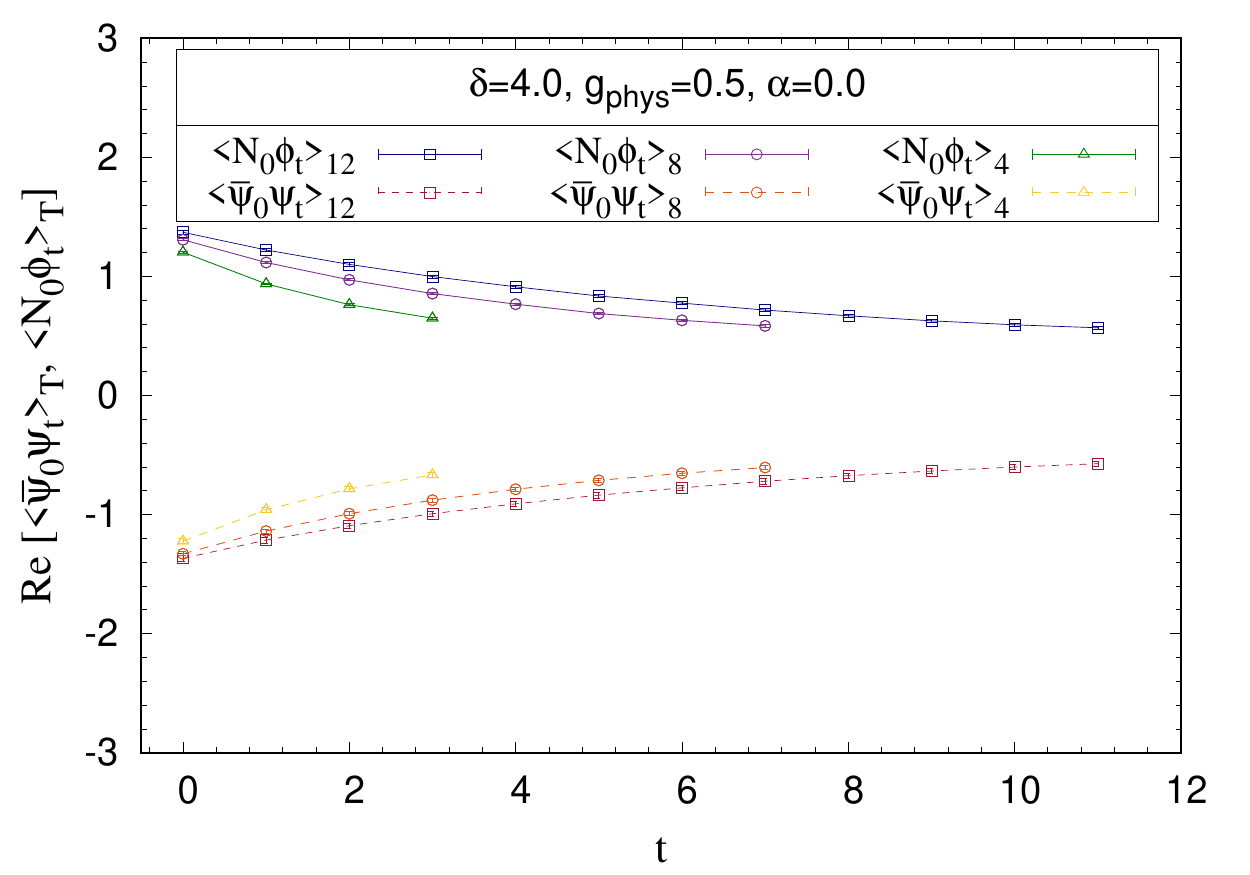}
	\includegraphics[width=.32\textwidth,origin=c,angle=0]{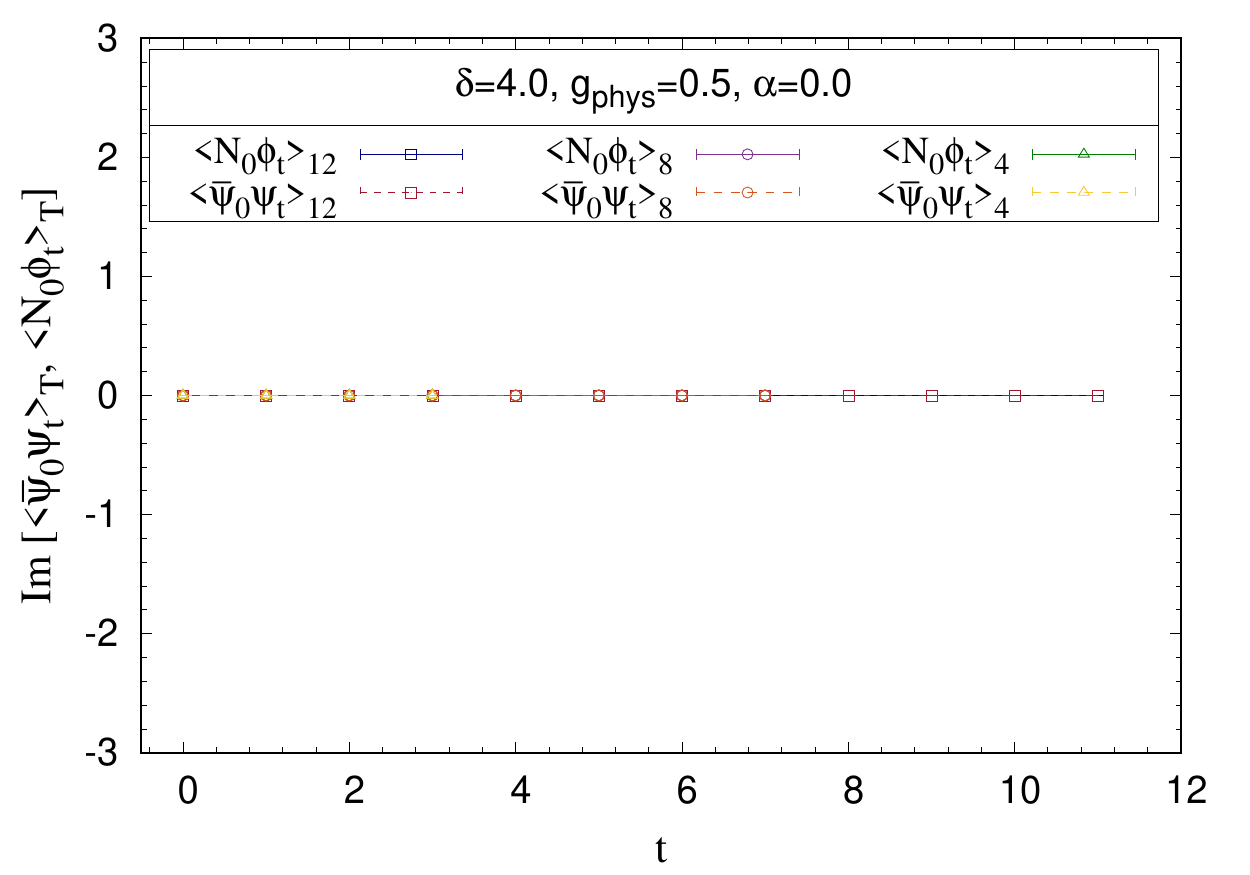}
	
	\caption {The  $\mathcal{PT}$-symmetric model with $\delta = 4$. The full Ward identity (left) and real (middle) and imaginary (right) parts of bosonic and fermionic contributions to Ward identity, for lattices with $T = 4, 8$, and $12$.}
	\label{fig:lat-susy-d4-T8-B-Sb}	
\end{figure*}

In Figs. \ref{fig:lat-susy-d2-T8-B-Sb} and \ref{fig:lat-susy-d4-T8-B-Sb} we show the Ward identities for $\delta = 2,~4$, respectively, on a $T = 8$ lattice. The full Ward identity $\mathscr{W}_1$ is shown on the left panel, and the real and imaginary parts of the bosonic and fermionic contributions to the Ward identity are shown in the middle and right panels. Our simulations show that the bosonic and fermionic contributions cancel each other out within statistical errors, and hence the Ward identities are satisfied. All these results clearly suggest that SUSY is preserved in models with $\mathcal{PT}$-symmetry inspired $\delta$-even potentials. 

\subsection*{Odd $\delta$ case}

In these models we had to introduce a deformation parameter to handle the singular drift problem. The results are extracted in the vanishing limit of this parameter.

The simulations were carried out for various non-zero $\mu$, and then the $\delta = 1$ model is recovered by taking ${\mu_{\rm phys} \to 0}$ limit. Our simulations suggest that when $\mu_{\rm phys}$ is above a particular value the correctness criteria is satisfied and the probability of absolute drift falls off exponentially. We take in account the $\mu_{\rm phys}$ parameter space where CL simulations are justified and consider the limit $\mu_{\rm phys} \to 0$. 

\begin{figure*}[tbp]
	\centering
	\includegraphics[width=.45\textwidth,origin=c,angle=0]{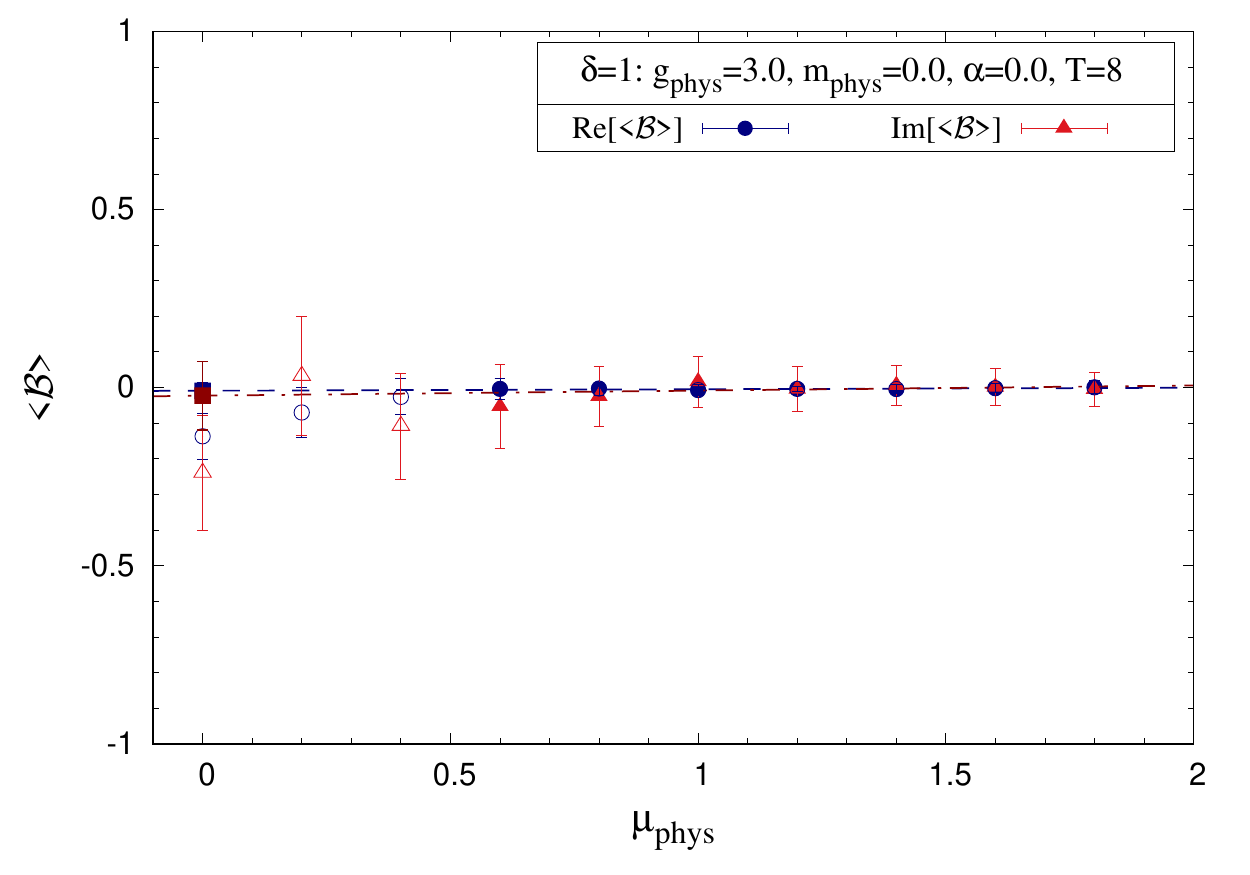} $~~~$ \includegraphics[width=.45\textwidth,origin=c,angle=0]{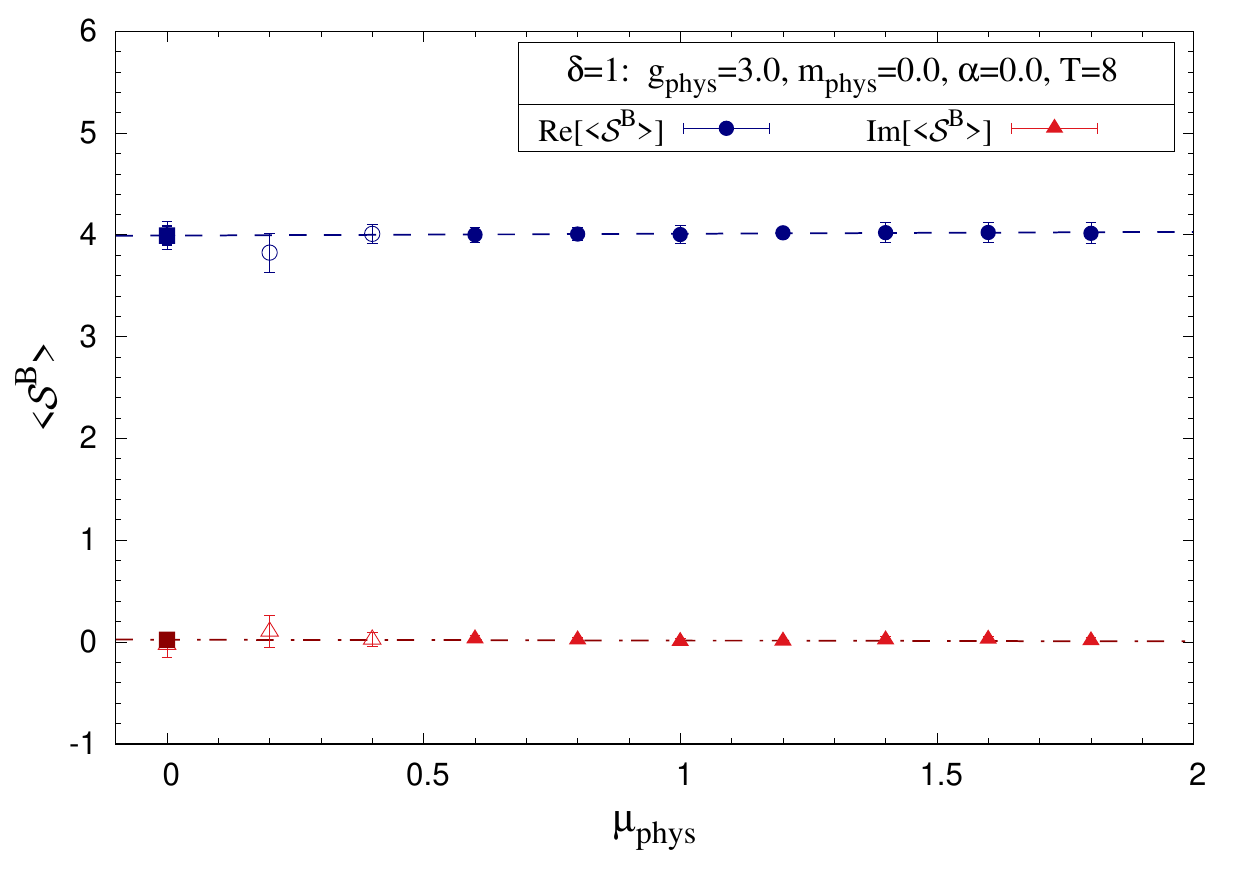}
	
	\caption{ The $\delta = 1$ model. Expectation values of $\B_\alpha$ (left) and $\mathcal{S}^B_\alpha$ (right) for various $\mu_{\rm phys}$ and $\lim \mu_{\rm phys} \to 0$ on a $T = 8$ lattice. Parameters used in the simulations are $g_{\rm phys} = 3$, $m_{\rm phys} = 0$, and $\alpha = 0$. }
	\label{fig:d1bSb}	
\end{figure*}

In Fig. \ref{fig:d1bSb}, we show the expectation values $\langle \B \rangle$ (left) and $\langle \mcS^{B} \rangle$ (right) for a lattice with $T = 8$ and for various $\mu_{\rm phys}$ values. The filled data points (red triangles for imaginary part and blue circles for real part) represent expectation values of observables for the parameter space where CL can be trusted, while for unfilled data points where the CL correctness criteria is not satisfied. Lines represent linear fit of observables for parameter space where CL simulations are justified and the solid squares represent values of respective observables in the $\lim \mu_{\rm phys} \to 0$. 

These simulation results indicate that $\langle \B  \rangle$ vanishes in the limit $\mu_{\rm phys} \to 0$. Also, the expectation value of the bosonic action $\langle \mcS^B \rangle = \hf T$ in this limit. It is also found to be independent of the physical parameters. Thus we conclude that SUSY is preserved in the $\delta = 1$ model. 

Now for the $\delta = 3$ model, inspired by the idea mentioned in Ref. \cite{Ito:2016efb} (which was successfully applied in Refs. \cite{Anagnostopoulos:2017gos, Anagnostopoulos:2020xai}) to handle the singular drift problem, we introduce a fermionic deformation term in the action. The fermionic action then becomes
\beq
\label{eqn:df}
\mcS^F = \sum_{i = 0}^{T-1} \psib_i  \bigg(\sum_{j = 0}^{T-1}  \nabla^{-}_{ij} + d_f + \Xi^{''}_{ij} \bigg) \psi_j,
\eeq
where $d_f$ is the deformation parameter.

The values of $d_f$ are chosen such that CL correctness criteria are satisfied. The $\delta = 3$ model is recovered as ${d_f \to 0}$. Our simulations suggest that above a particular $d_f$ value the correctness criteria is satisfied and the probability of absolute drift falls off exponentially. 

\begin{figure*}[tbp]
	\centering	
	\includegraphics[width=.45\textwidth,origin=c,angle=0]{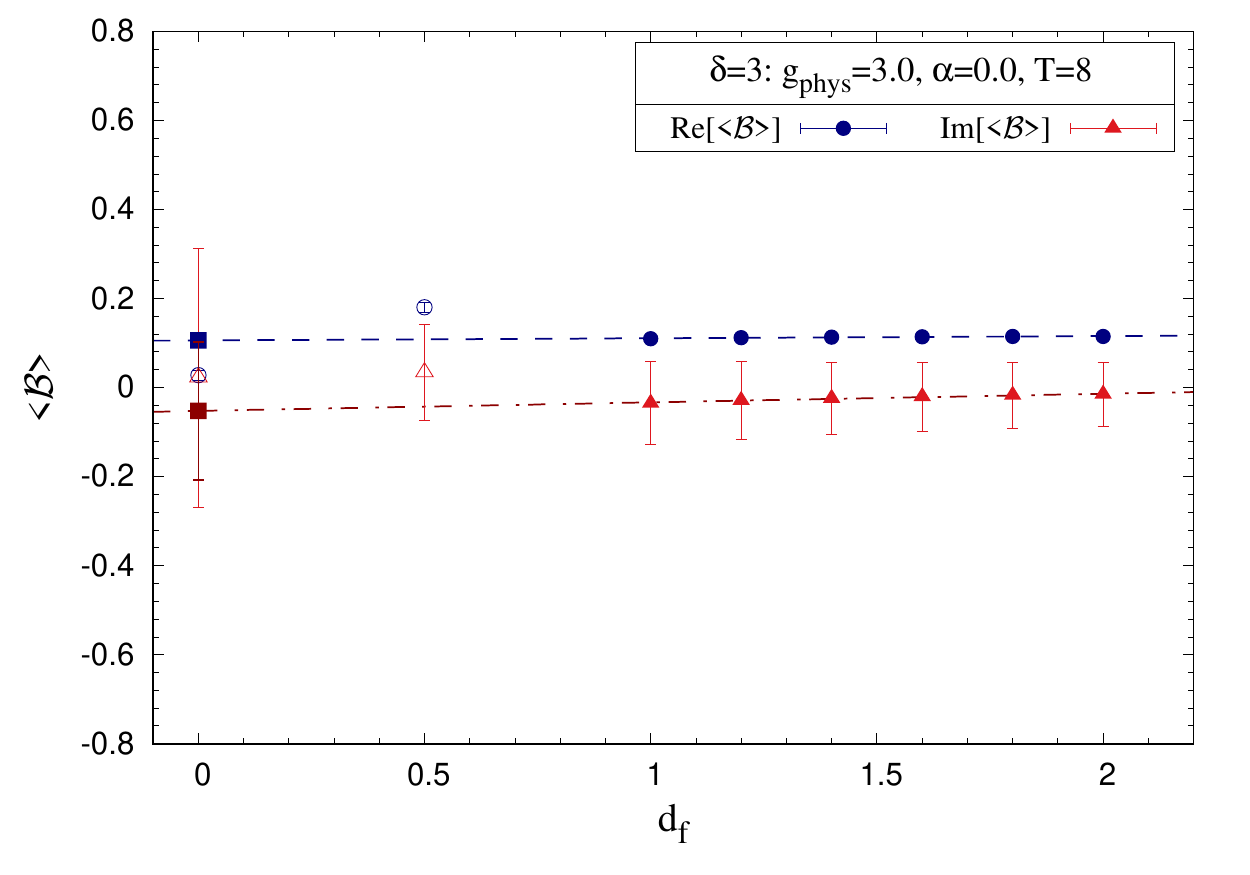}	
	\includegraphics[width=.45\textwidth,origin=c,angle=0]{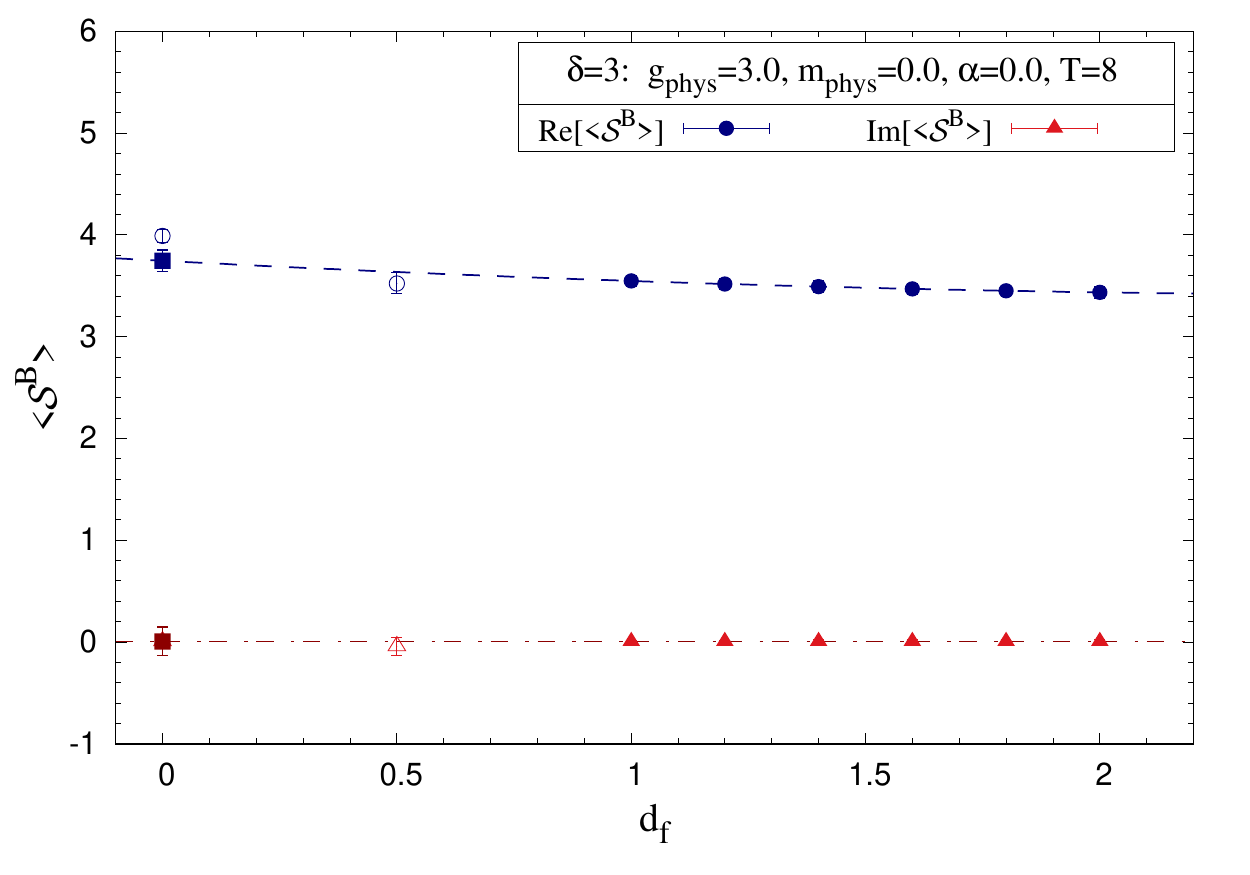}

	\caption{ The $\delta = 3$ model. Expectation values of $\B_\alpha$ (left) and $\mathcal{S}^B_\alpha$ (right) against mass deformation $d_f$ parameter on a $T = 8$ lattice. The parameters used are $g_{\rm phys} = 3$, $m_{\rm phys} = 0$, and $\alpha = 0.0$. The dashed curves represent extrapolations to the $d_f \to 0$ limit.}
	\label{fig:d3bSb}	
\end{figure*}

In Fig. \ref{fig:d3bSb}, we show $\langle \B \rangle$ (left) and $\langle \mcS^{B} \rangle$ (right) on a $T = 8$ lattice for various values of $d_f$. The filled data points (red triangles for the imaginary part and blue circles for the real part) represent the expectation values of the observables for the parameter space where CL can be trusted, while the corresponding unfilled data points represent the simulation data that do not satisfy the correctness criteria. The dashed curves represent a linear fit for $\langle \B \rangle$ data and quadratic fit for $\langle \mcS^{B} \rangle$ data. The solid squares represent the values of respective observables in the $\lim d_f \to 0$ limit. The simulation results suggest that $\langle \B  \rangle$ vanishes in the $\mu_{\rm phys} \to 0$ limit. Also, $\langle \mcS^B \rangle = \hf T$ within error bars in this limit. It is also found to be independent of the physical parameters of the model. These results indicate that SUSY is preserved in the $\delta = 3$ model. 

\section{Conclusions}
\label{sec:Conclusions}

In this paper, using complex Langevin method, we have investigated dynamical SUSY breaking in quantum mechanics models with real and complex actions. 

When periodic boundary conditions are used in quantum mechanics with broken SUSY the expectation values of observables are ill-defined. We resolved this problem by using twisted boundary conditions \cite{Kuroki:2009yg} in our non-perturbative lattice simulations.

We find that for the case of real actions, SUSY is preserved in models with harmonic and anharmonic oscillator potentials, and with odd-powered polynomial potential. SUSY is dynamically broken for the case of even-powered polynomial potential. For the case of supersymmetric anharmonic oscillator, our simulations reproduced the earlier results obtained through Hamiltonian Monte Carlo \cite{Catterall:2000rv}.

We then moved on to simulate an interesting class of supersymmetric models with complex actions that are ${\cal PT}$-symmetric. Our simulations suggested that SUSY is preserved in these models. 

We noticed that during the complex Langevin simulations some of these models suffered from the singular drift problem. In order to overcome this difficulty we introduced appropriate deformation parameters in the models, such that the criteria of correctness of complex Langevin simulations are respected. The target theories are recovered by taking the limits in which the deformation parameters go to zero. 

The reliability of the simulations were checked by studying the Langevin operator on observables and the exponential fall off of the drift terms. The results of our investigations on the correctness of the simulations are provided in Appendix \ref{subsec:reliability}. Our conclusion is that the complex Langevin method can be used reliably to probe non-perturbative SUSY breaking in various quantum mechanics models with real and complex actions. 

It would be interesting to extend our investigations to models in higher dimensions, especially quantum field theory systems in four dimensions, such as QCD with finite temperature and baryon/quark chemical potentials. Another long term hope would be to apply these methods to ${\cal PT}$-symmetric supersymmetric quantum field theories in higher dimensions. We hope that these studies may find applications in fundamental physics.

\acknowledgments

We thank discussions with Takehiro Azuma, Pallab Basu, and Navdeep Singh Dhindsa. The work of AJ was supported in part by the Start-up Research Grant (No. SRG/2019/002035) from the Science and Engineering Research Board (SERB), Government of India, and in part by a Seed Grant from the Indian Institute of Science Education and Research (IISER) Mohali. AK was partially supported by IISER Mohali and a CSIR Research Fellowship (Fellowship No. 517019).

\appendix

\section{Reliability of simulations}
\label{subsec:reliability}

In order to monitor the reliability of simulations we use the two recently proposed methods: one tracks the vanishing nature of the Fokker-Planck operator \cite{Aarts:2009uq, Aarts:2011ax, Aarts:2013uza} and the other monitors the decay of the probability distribution of the drift term magnitude \cite{Nagata:2016vkn, Nagata:2018net}. 

\subsection{Langevin operator on observables}
\label{app:FP-correctness}

The observables of the theory $\cO_i[\phi, \theta]$ at $i$-th site evolve in the following way 
\beq
\frac{\partial \cO_i[\phi, \theta]}{\partial \theta} = \widetilde{L}_i \cO_i[\phi, \theta], 
\eeq
where
$\widetilde{L}_i$ is the Langevin operator for the $i$-th site. It is defined as
\beq
\widetilde{L}_i = \left(\frac{\partial}{\partial \phi_i}  - \frac{\partial \mcS^{\rm eff}[\phi]}{\partial \phi_i}  \right) \ \frac{\partial}{\partial \phi_i}.
\eeq

Once the equilibrium distribution is reached, we can remove the $\theta$ dependence from the observables. Then we have $C_{\cO_i} \equiv \langle \widetilde{L}_i  \cO_i [\phi] \rangle = 0$, and this can be used as a criterion for correctness of the simulations. This criterion was used to study various interesting models in Refs. \cite{Aarts:2009uq, Aarts:2011ax, Aarts:2013uza}. 

If we take $\B_i$ at the $i$-th site as the observable then we have
\beq
\widetilde{L}_i \B_i =  -i \Xi^{'''}_{iii} + i \Xi^{''}_{ii} \frac{\partial \mcS^{\rm eff}}{\partial \phi_i}.
\eeq
Note that $\widetilde{L} \B$ respects translational symmetry on the lattice, and hence we can monitor the value obtained by averaging over all lattice sites. 

In table \ref{tab:LB-anho} we provide the expectation values of $\widetilde{L} \B$ for the simulations of supersymmetric anharmonic oscillator model discussed in Sec. \ref{subsec:susy-anho}. We see that this observable is zero within error bars and thus we can trust the simulations. 

\begin{table}[tbp]
	\begin{center}
	{\small
	\begin{tabular}{|c| l r |c|} 
		\hline
		$\Xi'(\phi)$&		$T$ &  $a = T^{-1}$   &  $~\langle \widetilde{L}\B_{\alpha}\rangle $  \\
		\hline
		\hline
		$ m \phi + g  \phi^3$
		&{8}
		&{0.125}& $0.0000(0) 	+ i0.0268(910)$ \\
		&{16}
		&{0.0625}&$0.0000(0) 	- i0.0379(450)$ \\
		&{32}
		&{0.03125}&$0.0000(0) 	+ i0.0131(232)$ \\
		&{64}
		&{0.015625}&$0.0000(0) 	+ i0.0056(132)$ \\ 
		\hline
	\end{tabular}
	}	
	\caption{Expectation value of $\widetilde{L} \B_\alpha$ for supersymmetric anharmonic oscillator with parameters $m_{\rm phys} = 10.0$ and $g_{\rm phys} = 100.0$. Simulations were performed for different lattice spacings with $\beta = 1$ and $\alpha = 0$.}
	\label{tab:LB-anho}
\end{center}
\end{table}

\begin{figure}[tbp]
	\centering
	\includegraphics[width=.49\textwidth,origin=c,angle=0]{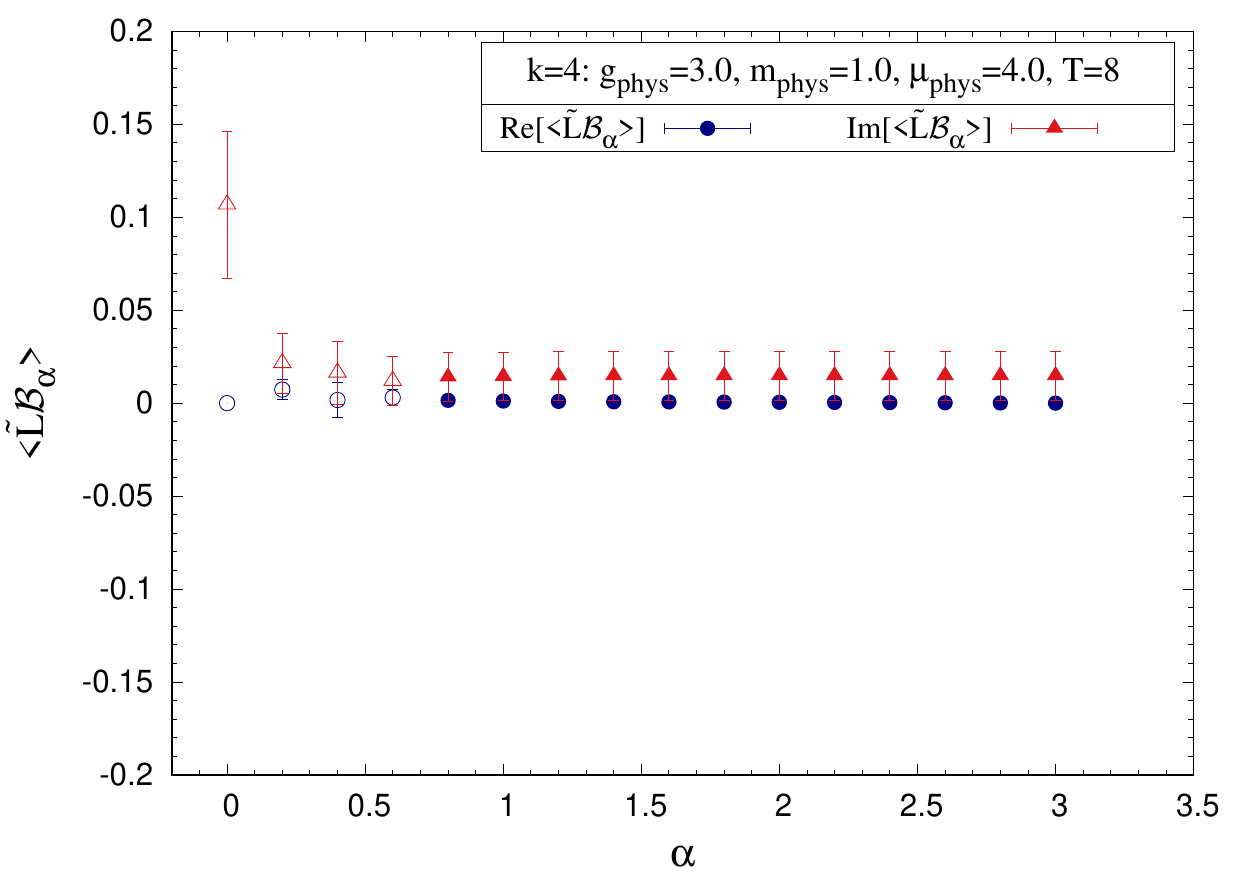}
	
	\caption {Expectation values of $\widetilde{L} \B_\alpha$ for various $\alpha$ values, on a $T = 8$ lattice, for the model with even-degree ($k = 4$) polynomial potential, given in Eq. \eqref{eqn:lat-gen-pot-k4}. The parameters used are $g_{\rm phys} = 3$, $m_{\rm phys} = 1$, and $\mu_{\rm phys} = 4.0$.}
	\label{fig:lbk4}
\end{figure}

In Fig. \ref{fig:lbk4}, we show the expectation values $\widetilde{L} \B$ for various twist parameter values for the model with case of real even-degree ($k = 4$) polynomial potential (discussed in Sec. \ref{subsec:gen-sps}). The filled data points (red triangles for imaginary part and blue circles for real part) represent expectation values of $\widetilde{L} \B$ for the parameter space where CL can be trusted, when the second criterion, decay-of-the-drift-term, to be discussed in the next section, is applied. The unfilled data points represent the expectation values of $\widetilde{L} \B$ in the parameter space where the decay-of-the-drift-term criterion is not satisfied.

In table \ref{tab:lbk5} we provide the expectation values of $\widetilde{L} \B$ for the simulations of real odd-degree ($k = 5$) polynomial potentials discussed in Sec. \ref{subsec:gen-sps}. We see that this observable is zero within error bars and thus we can trust the simulations.

\begin{table}[tbp]
	\begin{center}
		{\small
			\begin{tabular}{|c| l r |c|}  
				\hline
				$\Xi'(\phi)$&		$T$ &  $a = T^{-1}$  &  $~\langle \widetilde{L}\B_{\alpha}\rangle $ \\     	
				\hline
				\hline
				&{8}
				&{0.125}& $0.0000(0) 	- i0.0010(729)$ \\	
				${\Xi{'}}^{(5)}$
				&{12}
				&{0.0833}& $0.0000(0) 	+ i0.0127(508)$ \\	
				&{16}
				&{0.0625}& $0.0000(0) 	- i0.0301(396)$ \\ 	
				\hline
			\end{tabular}
		}
		\caption {Expectation values of $\widetilde{L} \B_\alpha$ for the model with odd-degree ($k = 5$) real-polynomial potential given in Eq. \eqref{eqn:lat-gen-pot-k5}. The parameters used are $g_{\rm phys} = 3$, $m_{\rm phys} = 1$, and $\mu_{\rm phys} = 4.0$.}
		\label{tab:lbk5}
	\end{center}
\end{table}

In table \ref{tab:lat-susy-pt-symm-d1-4-LB} we show the expectation values of $\widetilde{L} \B$ for $\mathcal{PT}$ symmetric models with $\delta = 2$ and 4. 

\begin{table}[tbp]
	\begin{center}
	{\small
	\begin{tabular}{|c| l r |c|}  
		\hline
		$\Xi'(\phi)$&$T$ &  $a=T^{-1}$   &  $\langle \widetilde{L}\B_{\alpha} \rangle$  \\   
		\hline
		\hline
		&${4}$&${0.25}$  & $-0.0000(0) - i0.0104(72)$ \\
		$\delta={2}$&${8}$&${0.125}$ & $-0.0000(0) + i0.0006(59)$ \\
		&${12}$&${0.0833}$  & $-0.0000(0) - i0.0104(72)$ \\
		\hline
		&${4}$&${0.25}$ & $0.0000(0) + i0.0403(244)$ \\
		$\delta={4}$&${8}$&${0.125}$  & $0.0000(0) + i0.0027(91)$ \\
		&${12}$&${0.0833}$  & $0.0000(0) - i0.0098(64)$ \\ 
		\hline
	\end{tabular}
	}
	\caption{Expectation value of $\widetilde{L}\B_\alpha$ for the $\mathcal{PT}$-symmetric potentials given in Eq. \eqref{eqn:lat-pt-symm-pot} with $\delta = 2$ and 4. The simulation parameters used are $\beta = 1$, $g_{\rm phys} = 0.5$, and $\alpha = 0$.}
	\label{tab:lat-susy-pt-symm-d1-4-LB}
	\end{center}
\end{table}

\begin{figure}[tbp]
	\centering
	\includegraphics[width=.45\textwidth,origin=c,angle=0]{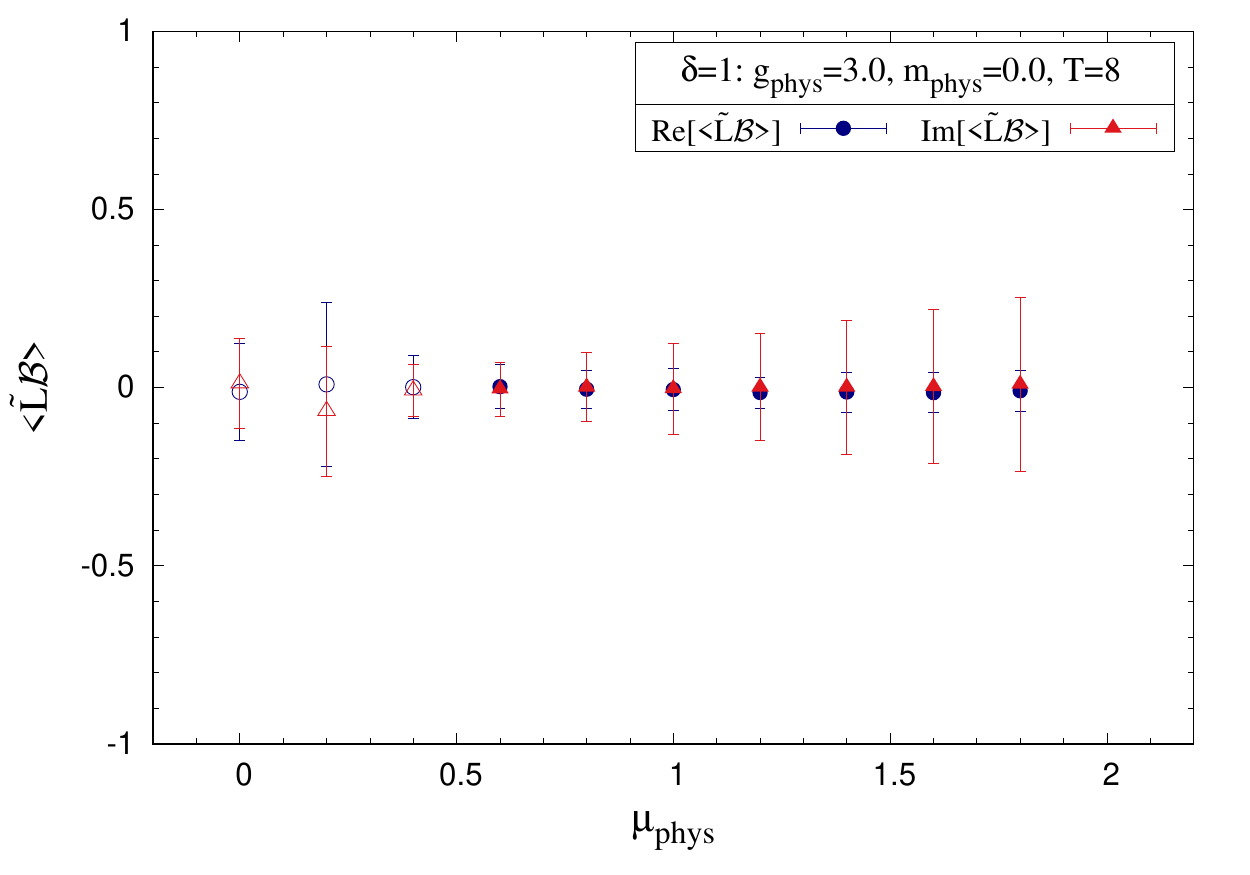} $~$ \includegraphics[width=.45\textwidth,origin=c,angle=0]{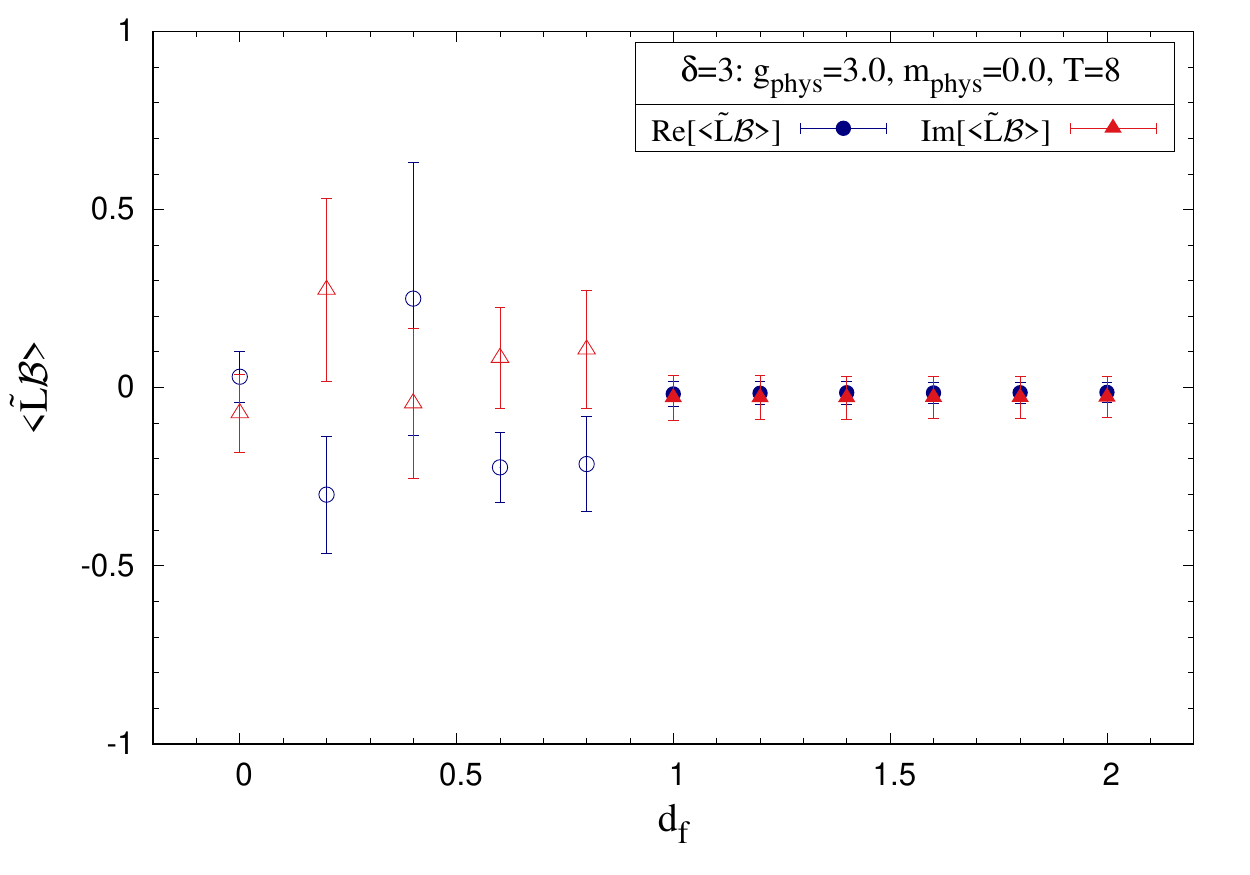}	
	
	\caption{Expectation values of $\widetilde{L} \B_\alpha$ for the $\mathcal{PT}$ symmetric model on a $T = 8$ lattice. (Left) Case $\delta = 1$. (Right) Case $\delta = 3$. The parameters used are $g_{\rm phys} = 3$, $m_{\rm phys} = 0$, and $\alpha = 0.0$.}
	\label{fig:lbd1_fig:lbd3}
\end{figure}

In Fig. \ref{fig:lbd1_fig:lbd3} (left), we show the expectation values for various $\mu_{\rm phys}$ values of $\widetilde{L} \B$ for the simulations of $\delta = 1$ model. The filled (unfilled) data points represent the simulations that does (does not) respect the decay-of-the-drift-term criterion. In Fig. \ref{fig:lbd1_fig:lbd3}, (right) we show $\widetilde{L} \B$ for various deformation parameter $d_f$ for the $\delta = 3$ model. Again, the filled (unfilled) data points represent the simulations that does (does not) respect the decay-of-the-drift-term criterion. 

\subsection{Decay of the drift terms}
\label{app:drift-decay}

The decay-of-the-drift-term criterion was proposed in Refs. \cite{Nagata:2016vkn, Nagata:2018net}. There, the authors demonstrated, in a few simple models, that the probability of the drift term should be suppressed exponentially or faster at larger magnitudes to guarantee the correctness of the CL method.

The magnitude of the mean drift is defined as
\beq
u \equiv \sqrt{\frac{1}{T} \sum_{i = 0}^{T-1} \left| \frac{\partial \mcS^{\text{eff}}}{\partial \phi_i}\right|^2}. 
\eeq

In our work, to avoid the singular drift problem, we introduced appropriate deformation parameters in the theory. The final results are obtained after extrapolating to the vanishing limits of deformation parameters.

\begin{figure}[tbp]
	\centering
	\includegraphics[width=.45\textwidth,origin=c,angle=0]{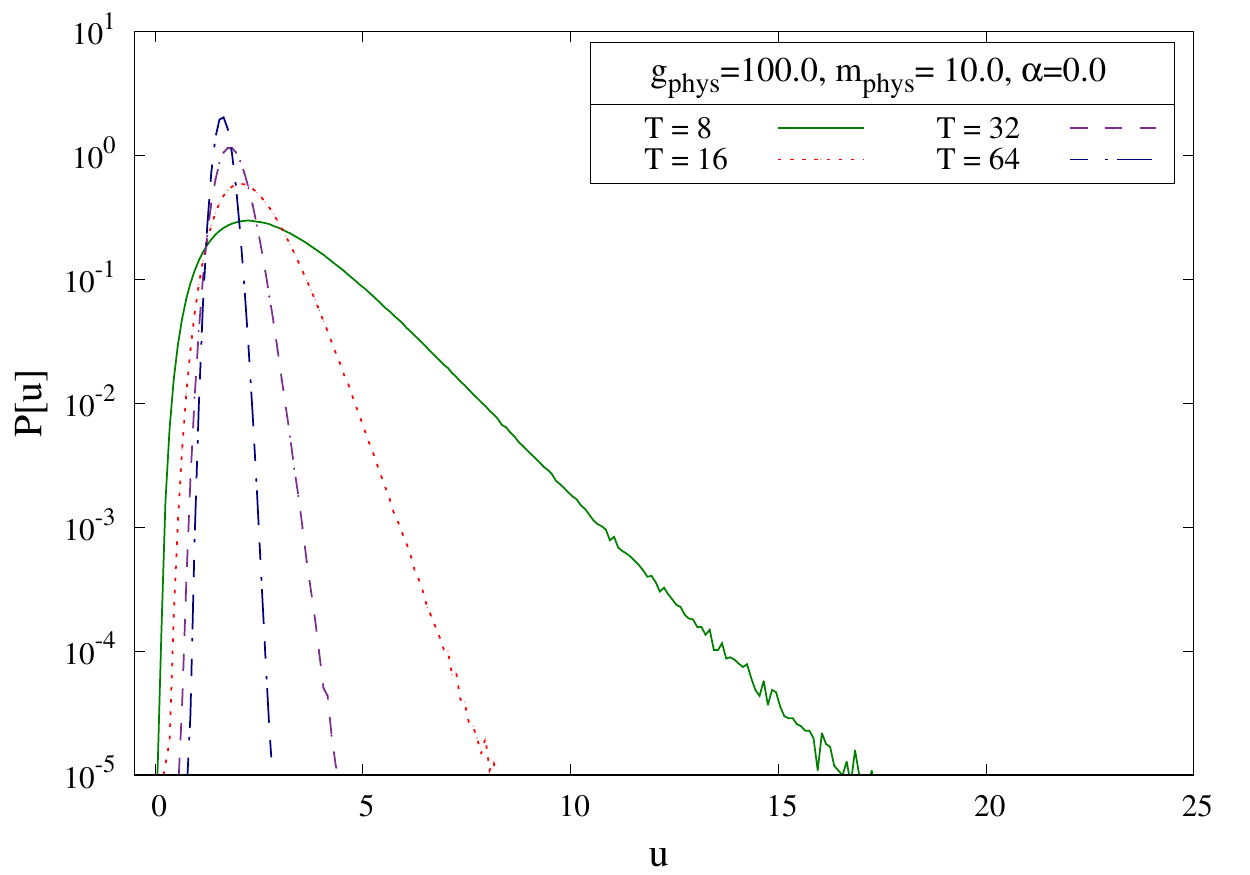}	$~$
	\includegraphics[width=.45\textwidth,origin=c,angle=0]{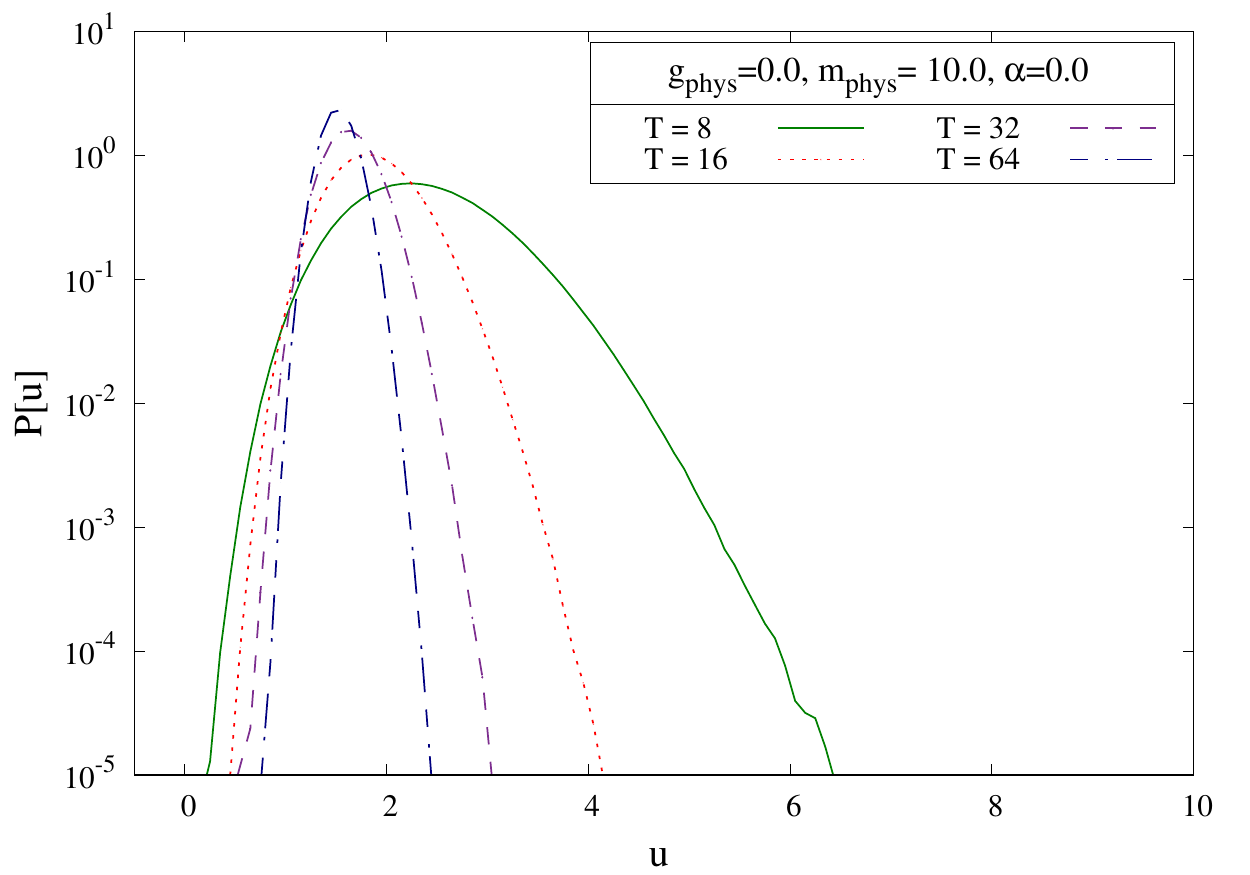}
	
	\caption{The decay of the drift terms. (Left) Supersymmetric anharmonic oscillator with the parameters $m_{\rm phys} = 10.0$ and $g_{\rm phys} = 100.0$. (Right) supersymmetric harmonic oscillator with parameters $m_{\rm phys} = 10.0$ and $g_{\rm phys} = 0.0$.}
	\label{fig:lat-anho-drift}
	
\end{figure}

In Fig. \ref{fig:lat-anho-drift}, we show the probability distribution $P(u)$ against $u$ for the simulations of supersymmetric anharmonic (left) and harmonic (right) potentials discussed in Sec. \ref{subsec:susy-anho}. We see that the drift terms decay exponentially or faster in these models and thus the simulations can be trusted.

\begin{figure}[tbp]
	\centering
	\includegraphics[width=.45\textwidth,origin=c,angle=0]{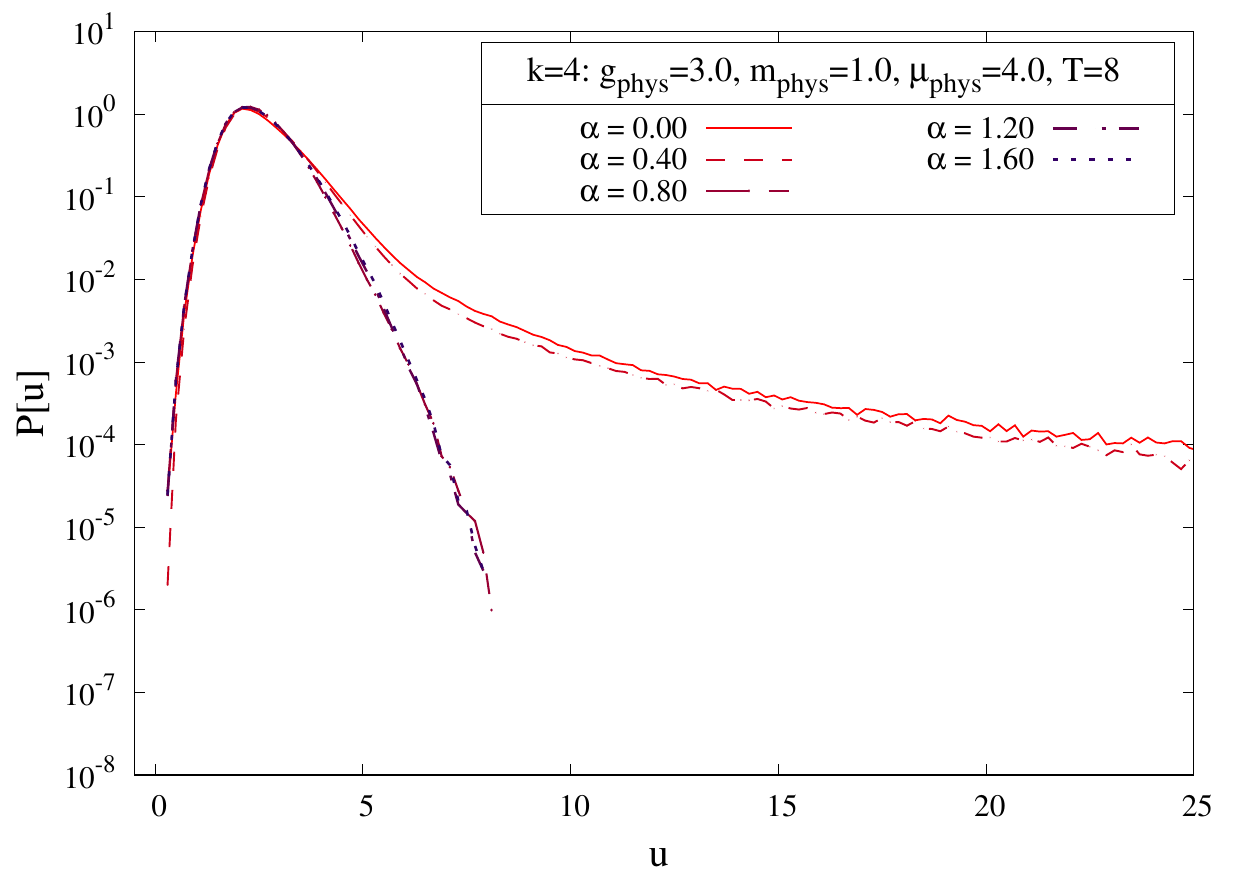}		$~$ \includegraphics[width=.45\textwidth,origin=c,angle=0]{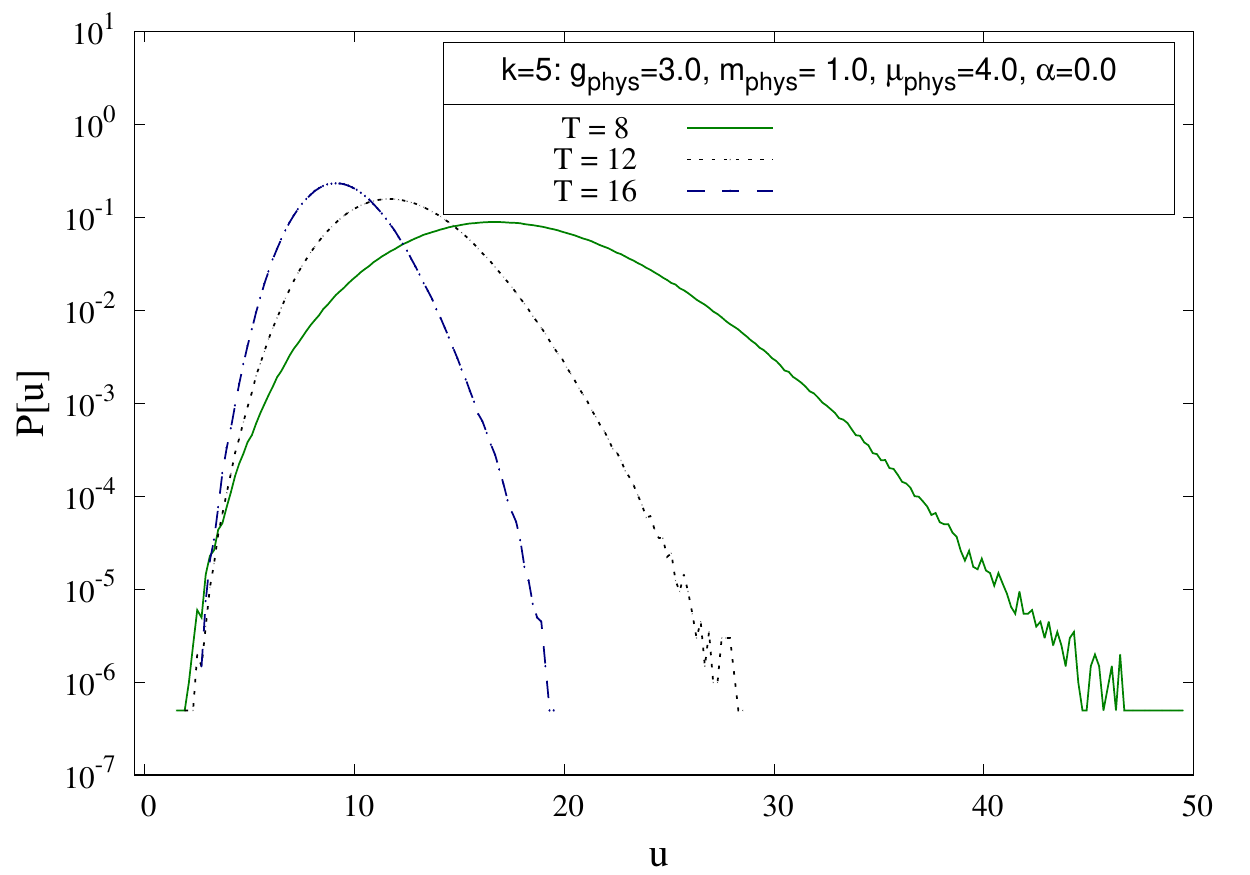}
	\caption {(Left) Decay of the drift term for various $\alpha$ values for the model with even-degree ($k = 4$) polynomial potential on a $T = 8$ lattice. The parameters used are $g_{\rm phys} = 3$, $m_{\rm phys} = 1$, and $\mu_{\rm phys} = 4.0$. (Right) The decay of the drift terms for the model with odd-degree ($k = 5$) real polynomial potential on $T = 8,12,16$ lattices. The parameters used are $g_{\rm phys} = 3$, $m_{\rm phys} = 1$, and $\mu_{\rm phys} = 4.0$.}
	\label{fig:pdSk4_fig:pdSk5}
	
\end{figure}

In Fig. \ref{fig:pdSk4_fig:pdSk5} (left) we show the decay of drift terms for various $\alpha$ values, for the model with even-degree ($k = 4$) real polynomial potential. We see that the drift terms decay exponentially or faster when $\alpha \geq 0.8$ and the simulations can be trusted in this parameter regime. In Fig. \ref{fig:pdSk4_fig:pdSk5} (right) we show the decay of drift terms for various $\alpha$ values, for the model with even-degree ($k = 5$) real polynomial potential. We see that the drift terms decay exponentially or faster when $\alpha = 0$ and the simulations can be trusted in this parameter regime.

\begin{figure}[tbp]
	\centering
	\includegraphics[width=.45\textwidth,origin=c,angle=0]{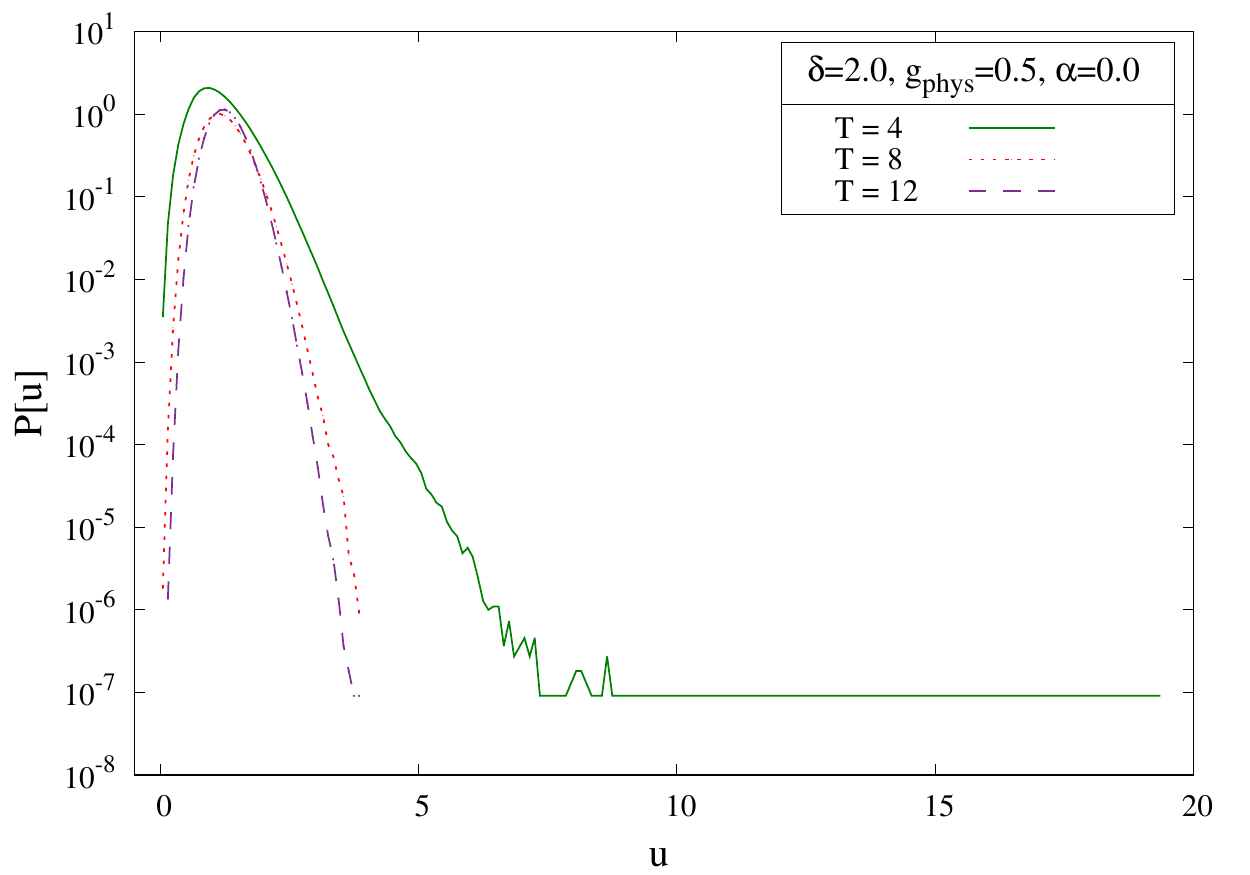}	$~$ \includegraphics[width=.45\textwidth,origin=c,angle=0]{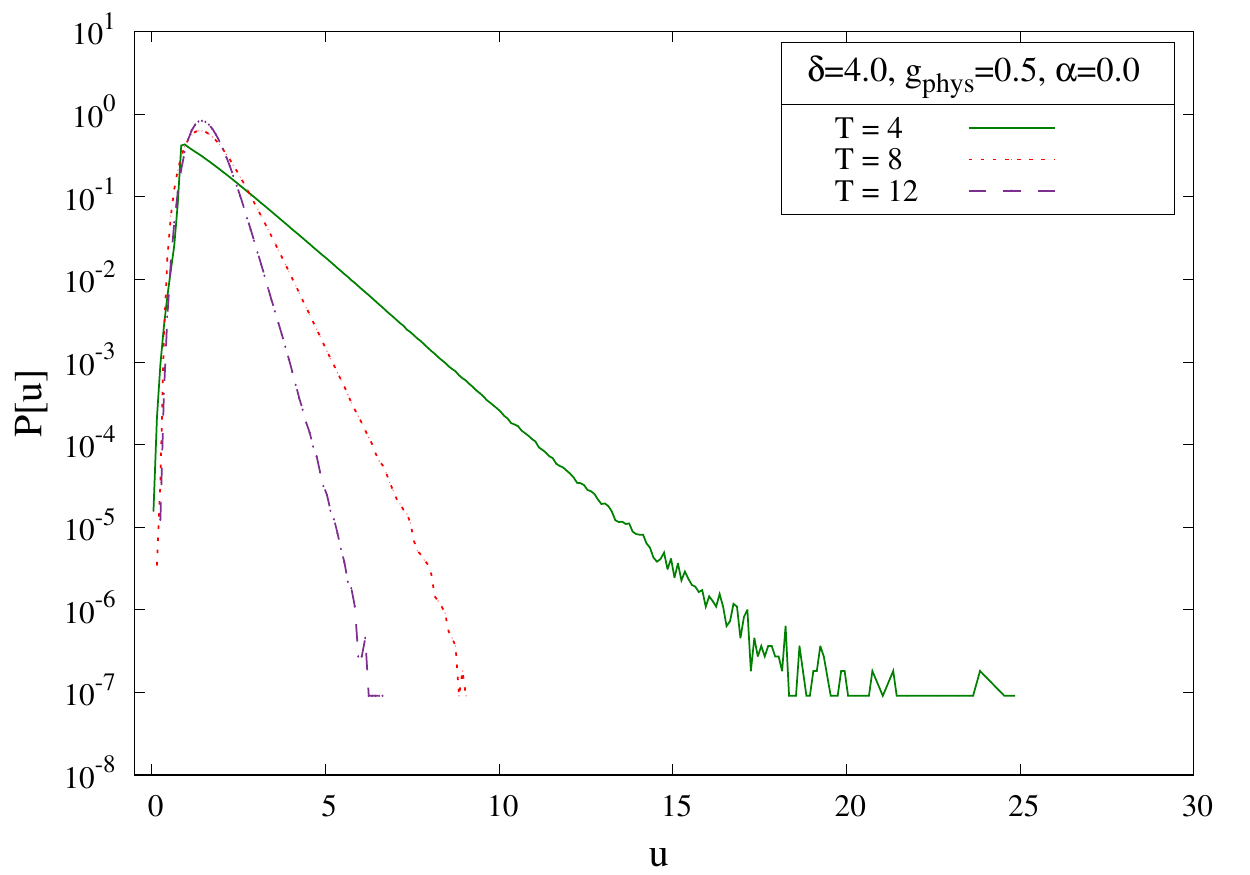}	
	\caption{Decay of the drift terms for the $\mathcal{PT}$ symmetric models with $\delta =  2$ (left) and $\delta =  4$ (right). The parameters used are $g_{\rm phys} = 0.5$ and $\alpha = 0$.}
	\label{fig:pdSd2-4}
\end{figure}

In Fig. \ref{fig:pdSd2-4} we show The drift term decay for the $\mathcal{PT}$ symmetric models with $\delta = 2$ (left) and $\delta = 4$ (right). The drift terms decay exponentially or faster when $\alpha = 0$. Thus the simulations can be trusted in this parameter regime.

\begin{figure}[tbp]
	\centering	
	\includegraphics[width=.45\textwidth,origin=c,angle=0]{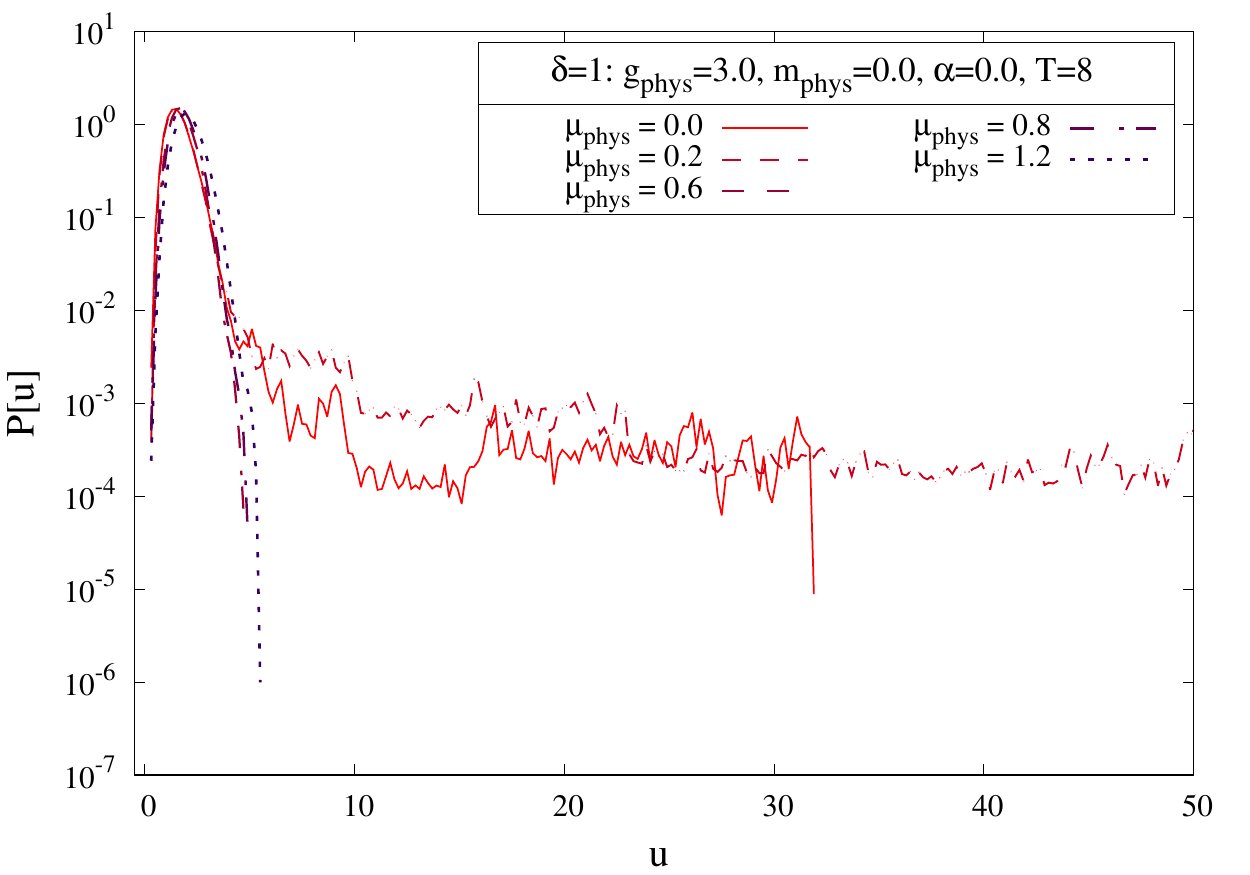} 	$~$ \includegraphics[width=.45\textwidth,origin=c,angle=0]{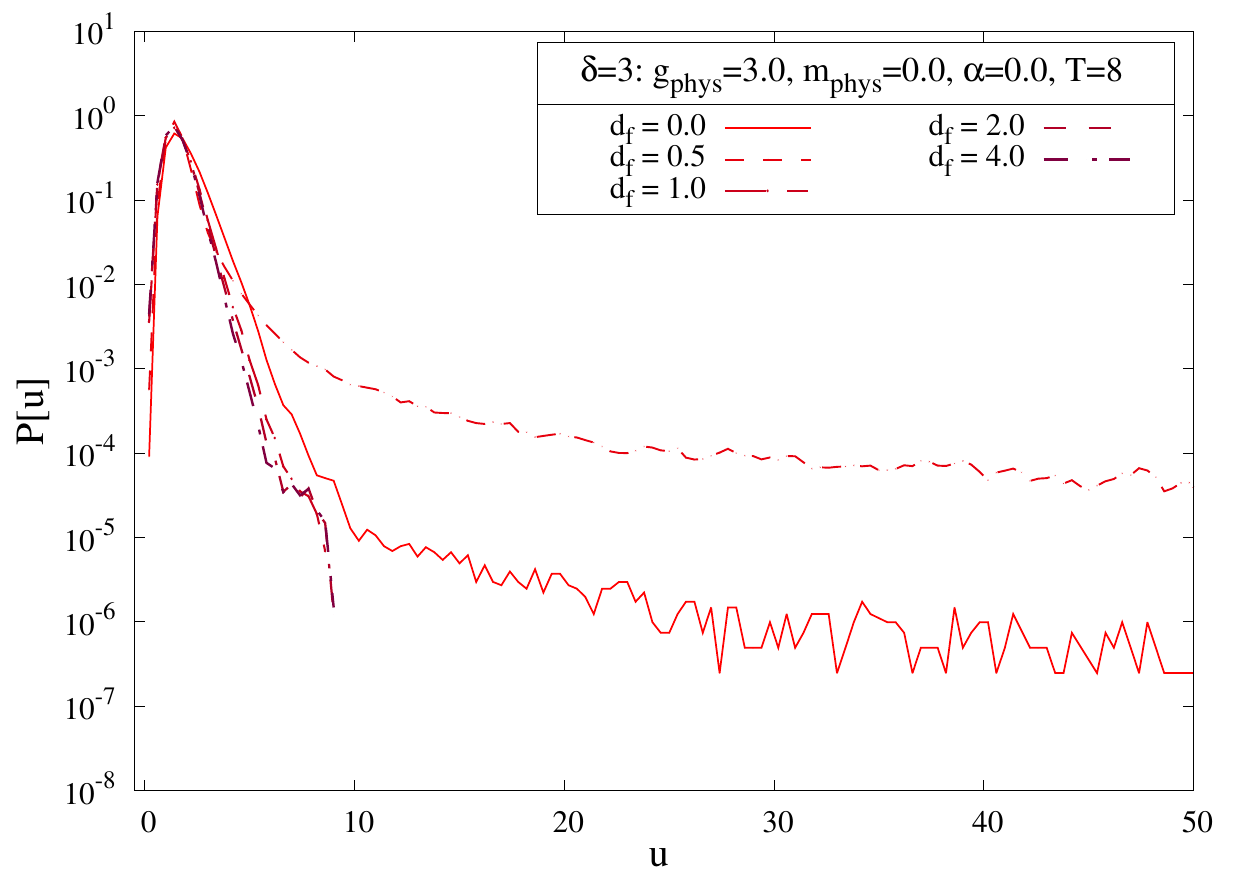}
	\caption{The decay of the drift terms for the $\mathcal{PT}$ symmetric models with odd $\delta$ on $T=8$ lattice. The parameters used were $g_{\rm phys} = 3$, $m_{\rm phys} = 0$, and $\alpha = 0.0$. (Left)  The $\delta = 1$ case. Simulations were performed for various $\mu_{\rm phys}$ values. (Right) The $\delta = 3$ case. Simulations were performed for various values of the fermionic mass deformation parameter $d_f$.}
	\label{fig:pdSd1_fig:pdSd3}
\end{figure}

In Fig. \ref{fig:pdSd1_fig:pdSd3} (left) we show the drift term decay for the $\mathcal{PT}$ symmetric model with $\delta = 1$, for various $\mu_{\rm phys}$ values. The drift terms decay exponentially or faster when $\mu_{\rm phys} \geq 0.6$. Thus the simulations can be trusted in this parameter regime. In Fig. \ref{fig:pdSd1_fig:pdSd3} (right) we show the decay of the drift terms for various values of the fermionic mass deformation parameter $d_f$, for the $\mathcal{PT}$ symmetric model with $\delta = 3$. Here we see that the drift terms decay exponentially or faster when $d_f > 1.0$ and thus the simulations can be trusted in this parameter regime.

\bibliography{paper.bib}
\bibliographystyle{JHEP.bst}

\end{document}